\documentclass[rivista,cite]{cimento}

\def\sens{{\cal S}}
\def\var{\hbox{var}}
\def\coh{{\rm coh}}
\def\TWB{{\rm TWB}}
\def\NRF{{\rm NRF}}
\def\Tr{{\rm Tr}}
\def\e{e}
\def\hH{{\cal H}}
\def\Ubs{U_{\rm BS}}
\def\ha{a}
\def\hb{b}
\def\hJ{J}
\def\hA{A}
\def\hB{B}
\def\hD{{\cal D}}
\def\hS{S}
\def\hrho{\varrho}
\def\hK{\hat{K}}
\newcommand{\rmpl}[1]{{\rm #1}}
\newcommand{\ket}[1]{\vert #1 \rangle}
\newcommand{\bra}[1]{\langle #1 \vert}
\newcommand{\braket}[2]{\langle #1 \vert #2 \rangle}

\usepackage{graphicx,bbm,amsmath}  

\title{High-precision innovative sensing\\ with continuous-variable optical states}
\author{Stefano~Olivares\from{DIP}\from{INFN}\thanks{E-mail: \texttt{stefano.olivares@fisica.unimi.it}}}
\instlist{
\inst{DIP} Quantum Technology Lab, Dipartimento di Fisica ``Aldo Pontremoli'',\\
Universit\`a degli Studi di Milano, via Celoria 16, I-20133 Milano, Italy
\inst{INFN} INFN, Sezione di Milano, via Celoria 16, I-20133 Milano, Italy}

\begin{document}

\maketitle

\begin{abstract}
When applied to practical problems, the very laws of quantum mechanics can
provide a unique resource to beat the limits imposed by classical physics: this is the
case of quantum metrology and high-precision sensing.
Here we review the main results obtained in the recent years at the Quantum Technology Lab
of the Department of Physics ``Aldo Pontremoli'' of the University of Milan, also in collaboration with
national and international institutions. In particular we focus on the application of
continuous-variable optical quantum
states and operations to improve different field of research, ranging from interferometry 
to more fundamental problems, such as the testing of quantum gravity.
\end{abstract}

\section{Introduction}

Quantum states of the radiation field represent a key ingredient to beat the limits
imposed by classical light. The possibility to generate and manipulate single-photon states
\cite{kwiat,cialdi:manip} has made possible the implementation of protocols beyond the possibility of
classical physical systems, paving the way to quantum cryptography \cite{QKD:1},
quantum communication \cite{COMM:1}, quantum enhanced sensing and metrology
\cite{QM,multipar:18} and the study of complex problems as, for instance, the boson sampling
\cite{boson:sci,boson:walm}.
When single photons are considered, the information is usually encoded into
observables with a discrete spectrum, such as the polarisation degree of freedom.
Unfortunately, the information encoded in single photon states may be completely
destroyed by the presence of losses: once the photon is lost, the carried information
cannot be retrieved, though one can resort to schemes more tolerant to dissipation \cite{din:14}.
\par
Therefore, many efforts have been made
to extend discrete-variable quantum information protocols to the continuous-variable analogue,
where the information is now
encoded into observables with continuous spectrum (but not only), such as the amplitude or
the phase of the field \cite{rev:gauss,rev:QI}. Continuous-variable optical states are usually more robust
to losses with respect to the single photons \cite{fading:12,pino:15}, since they may contain
mesoscopic \cite{ale:13} or macroscopic number of photons \cite{use:15,silb:02}.
\par
In this scenario, optical correlations, both quantum and classical, represent a fundamental resource for developing technologies, opening unprecedented opportunities in the fields of metrology, positioning, imaging and sensing. Furthermore, the correlations existing between two or more \cite{tri:olipa} light beams have also a theoretical interest, being of key relevance in quantum optics and quantum electrodynamics and are at the basis of the quantum information processing.
\par
In this paper we are going to summarise some of the main theoretical and experimental results we
achieved at the Quantum Technology Lab of the Department of Physics ``Aldo Pontremoli'' of
University of Milan also in collaboration with national and international teams. We will
focus on our research on optical continuous-variable states and their application to interferometry,
imaging and also to test more fundamental theories, such as the noncommutativity of position and
momentum at the Planck scale.
\par
The structure of the paper is as follows. Section~\ref{s:QET} introduces the basic tools of the estimation and
quantum estimation theory and, in particular, the role of the Fisher and quantum Fisher information.
\par
In section~\ref{s:QINT} we will see how it is possible to enhance the sensitivity of an interferometer
exploiting the quantum features of light. We present the main results we obtained when
continuous-variable states, such as coherent and squeezed states of the optical field, are used
to feed the interferometers: this will be the subject of section~\ref{s:bounds}, where we will study the
bound imposed by quantum mechanics to the precision of an interferometer when only
continuous-variable optical states are considered.
\par
Quantum correlations can be exploited to improve the detection. In section~\ref{s:IMAG:qill}
we present the first realisation of a practical quantum illumination protocol. It is a
scheme to detect the presence of a faint object embedded in a noisy background, exploiting
the quantum correlation existing between two light beams.
\par
The possibility to enhance the sensitivity of optical interferometers by using setups based on
two interferometers instead of a single one,
is discussed in sections~\ref{s:coupled:qgrav}, \ref{s:probing} and \ref{s:sq:corr:int}.
In section~\ref{s:probing} we show how two correlated interferometers can be
used to test quantum gravity and, more precisely, the noncommutativity of position and momentum
at the Planck scale, outperforming schemes based on classical light. Section~\ref{s:sq:corr:int}
investigates the role of a nonclassical resource, such as squeezing, to improve the performance of
correlated interferometry. Finally, we draw some concluding remarks in section~\ref{s:outro}.
\par
Throughout the manuscript we will use many concepts from quantum optics. The interested reader
can find a brief review of these basic notions in dedicated appendices.


\section{Basics of quantum estimation theory}\label{s:QET}

In this section we introduce the basic tools of estimation theory and quantum estimation theory
which we will use throughout this paper. In particular, we will introduce the concepts of
Fisher information and of quantum Fisher information.
\par
When a physical parameter is not directly accessible, one needs to resort to indirect measurements.
This is the case, for instance, of the measurement of a phase shift, of the entropy or of
the entanglement between two or more parties of a quantum system. Here we focus on the
single parameter estimation but our results can be extended to scenarios involving
more than one parameter \cite{paris:rev}.
\par
\begin{figure}[tb]
\begin{center}
\includegraphics[width=0.65\textwidth]{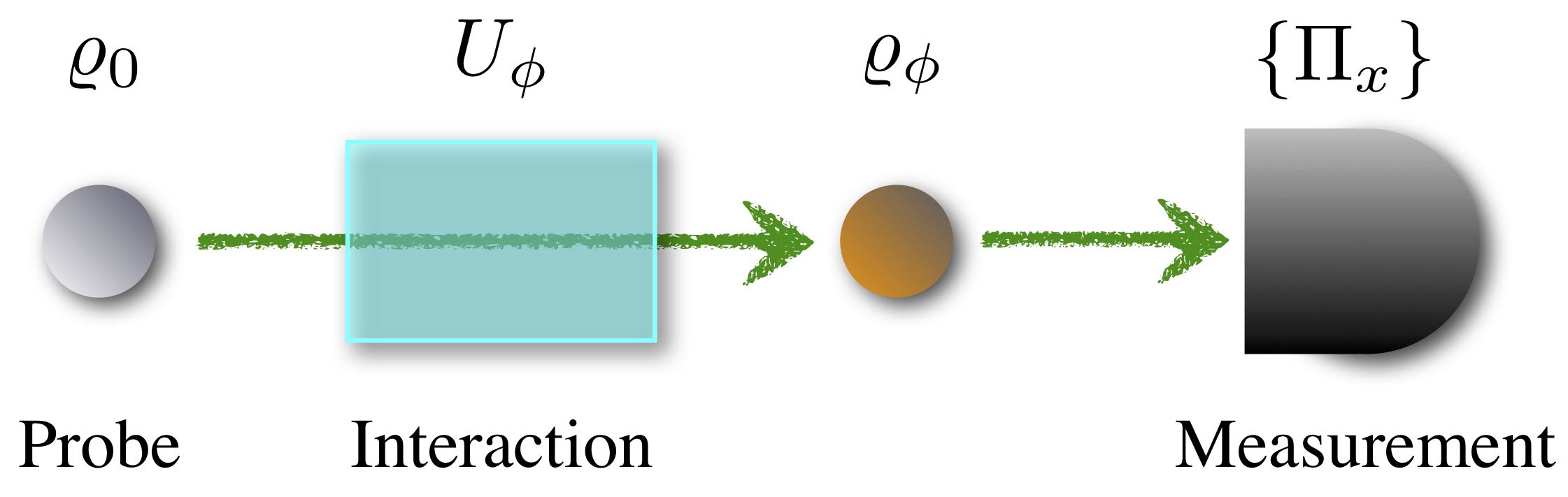}
\end{center}
\vspace{-0.3cm}
\caption{Scheme of a quantum estimation protocol: a quantum probe is prepared
in the initial state $\varrho_0$ and interacts with a system. The interaction, described by the
unitary operator $U_\phi$, is characterised by
the parameter $\phi$ we want to estimate. The evolved state is measured and the
outcomes are processed to retrieve the information about $\phi$.
In general the measurement is described by a positive operator-valued measure (POVM) $\{\Pi_x\}$, where $x$ is the outcome. See the text for details.}\label{f:QET:scheme}
\end{figure}
In fig.~\ref{f:QET:scheme} we sketch a typical setup to estimate one parameter, say $\phi$, through
a quantum probe. In a quantum estimation protocol, a quantum probe is prepared in
a known state described by the density operator $\varrho_0$. Then the probe interacts with a system,
the interaction being described by the unitary operator $U_\phi$. The evolved state of the probe,
$\varrho_\phi = U_\phi \varrho_0 U_\phi^\dag$, encodes the unknown parameter $\phi$
and, then, undergoes a measurement described in general by the positive-operator valued measure
(POVM) $\{ \Pi_x \}$, $x$ being the measurement outcome.
The data sample $\Omega = \{x_1, x_2, \ldots , x_M \}$ is finally processed to retrieve the
value of the parameter by means of a suitable estimator $\hat{\phi}$. Of course,
we are not only interested in the expectation value $E[\hat{\phi}]$ of $\hat{\phi}$,\
but also in its uncertainty $\var(\hat{\phi}) = E[\hat{\phi}^2] -  E[\hat{\phi}]^2$.
Therefore, the main goals of quantum estimation theory are to find the \emph{optimal probes}
which minimise the uncertainty and the \emph{optimal measurements} which allow to reach such
minimum uncertainty.
\par
From the classical point of view, optimal unbiased estimators saturate the Cram\'er-Rao inequality
or bound \cite{helstrom}:
\begin{equation}
 \var(\hat{\phi}) \ge \frac{1}{M F(\phi)}\,,
\end{equation}
and we introduced the \emph{Fisher information}:
\begin{equation}
F(\phi) = \sum_{k=1}^M p(x_k|\phi)\, [ \partial_\phi \log p(x_k|\phi)]^2\,,
\end{equation}
$p(x|\phi)$ being the conditional probability of the outcome $x$ given $\phi$.
\par
In the presence of a quantum probe and of a POVM, the conditional probability reads:
\begin{equation}
p(x|\phi) = \Tr[\varrho_\phi \Pi_x]\,,
\end{equation}
and, thus, the Fisher information can be written as (to be more general, we assume that $x \in \Omega
\subset {\mathbbm R}$):
\begin{eqnletter}
\label{Fisher:SLD}
F(\phi) &=& \int_\Omega
\frac{\Tr[\partial_\phi \varrho_\phi \Pi_x]^2}{\Tr[\varrho_\phi \Pi_x]}\,dx \,,\\[1ex]
&=& \int_\Omega
\frac{\Re{\rm e}\left\{\Tr[\varrho_\phi \Pi_x L_\phi]\right\}^2}{\Tr[\varrho_\phi \Pi_x]}\, dx \,,
\end{eqnletter}
where $L_\phi$ is the symmetric logarithmic derivative such that
\begin{equation}
\frac{\partial \varrho_\phi}{\partial \phi} =
\frac{L_\phi \varrho_\phi + \varrho_\phi L_\phi}{2}\,.
\end{equation}
We can now maximise $F(\phi)$ over all the possible POVMs to obtain the quantum Cram\'er-Rao bound
\cite{QCR:94,QCR:96}:
\begin{equation}
 \var(\hat{\phi}) \ge \frac{1}{M H(\phi)}\,,
\end{equation}
where $H(\phi) = \Tr[\varrho_\phi L_\phi^2] \ge F(\phi)$ is the \emph{quantum Fisher information}
\cite{helstrom,QFI}.
According to its definition, the eigenvectors of the symmetric logarithmic derivative $L_\phi$
correspond to the optimal POVM.
\par
Looking for the analytical expression of the POVM and the optimal measurement is not always
a simple task, since it requires an optimisation procedure. However, there are many cases in
which it is possible to find some explicit relations.
\par
In the simplest scenario the parameter
to be estimated is the amplitude of a unitary perturbation applied to the probe state $\varrho_0$.
This is the case, for instance, of a phase shift imposed to an optical field.
If ${\cal G}$ is the Hermitian generator of the perturbation, we can write $U_\phi = \exp(-i\, \phi\, {\cal G})$.
Now, expanding the input state in its eigenbasis, namely:
\begin{equation}
\varrho_0 = \sum_n \varrho_n | \psi_n \rangle \langle \psi_n |\,,
\end{equation}
we obtain  $\varrho_\phi = \sum_n \varrho_n | \psi_n^{(\phi)} \rangle \langle \psi_n^{(\phi)} |$,
with $| \psi_n^{(\phi)} \rangle = U_\phi | \psi_n \rangle$. It is straightforward to show that \cite{paris:rev}:
\begin{equation}
\label{QFI:SP}
H = 2 \sum_{n \ne m} \frac{(\varrho_n - \varrho_m)^2}{\varrho_n + \varrho_m}
| \langle \psi_n | {\cal G} | \psi_m  \rangle |^2\,,
\end{equation}
and it is independent of $\phi$. Moreover, if the input state is a pure state,
{\it i.e.} $\varrho_0 = | \psi_0 \rangle \langle \psi_0 |$, eq.~(\ref{QFI:SP}) reduces to:
\begin{eqnletter}
H &=&
4 \left( \langle \psi_0 | {\cal G}^2 | \psi_0  \rangle - \langle \psi_0 | {\cal G} | \psi_0  \rangle^2 \right) \,,\\[1ex]
&=& 4 \, \var[{\cal G}]\,,
\end{eqnletter}
namely, the quantum Fisher information is proportional to the fluctuations of the generator
${\cal G}$ on the probe state.
\par
In section~\ref{s:QINT} we will apply these results to find the
ultimate bounds to interferometric sensitivity. The interested reader can find
in refs.~\cite{paris:rev} and \cite{datta:rev} further details about the application
of quantum estimation theory to more general scenarios of interest for quantum information processing.
In particular, at our Quantum Technology Lab we applied the tools of quantum estimation
to quantum optics \cite{brunelli:12,bina:dike,rossi:17}, to open quantum systems \cite{geno:phase,geno:phase:exp,bina:env} and to more fundamental problems
\cite{tama:16,rossi:16,seveso:17}, just to cite some of the relevant fields of research we
investigated  in the last years.


\section{Quantum interferometry with continuous-variable states}\label{s:QINT}

In this section we introduce the reader to optical quantum interferometry
using the sensitivity of a typical interferometer and its connection with the Fisher information.
The application of quantum estimation theory to find the ultimate bounds given by the quantum
Fisher information and the analysis of more sophisticated setups, involving
also active elements, will be discussed in section~\ref{s:bounds}.
\par
An optical interferometer is a paradigmatic example of one of the most precise devices
available in physics. Its applications range from technological ones,
to the challenging task of modern cosmology, {\it i.e.} the direct detection of gravitational
waves, and also to measure Planck-scale effects predicted by quantum gravity theories.
An interferometer should maximise the precision in the estimation of phase-shift fluctuations
given some energy constraints on the energy circulating in the interferometer itself.
To this aim one can exploit the nonclassical features of the probe beams which
may outperform the performance of the corresponding classical ones, thus leaving
room for quantum enhanced interferometry. However, the fluctuations associated
with the very quantum nature of light pose bounds to the precision, which can be assessed
by the modern tools of quantum estimation theory briefly introduced in section~\ref{s:QET}.
\par
\begin{figure}[h!]
\begin{center}
\includegraphics[width=0.65\textwidth]{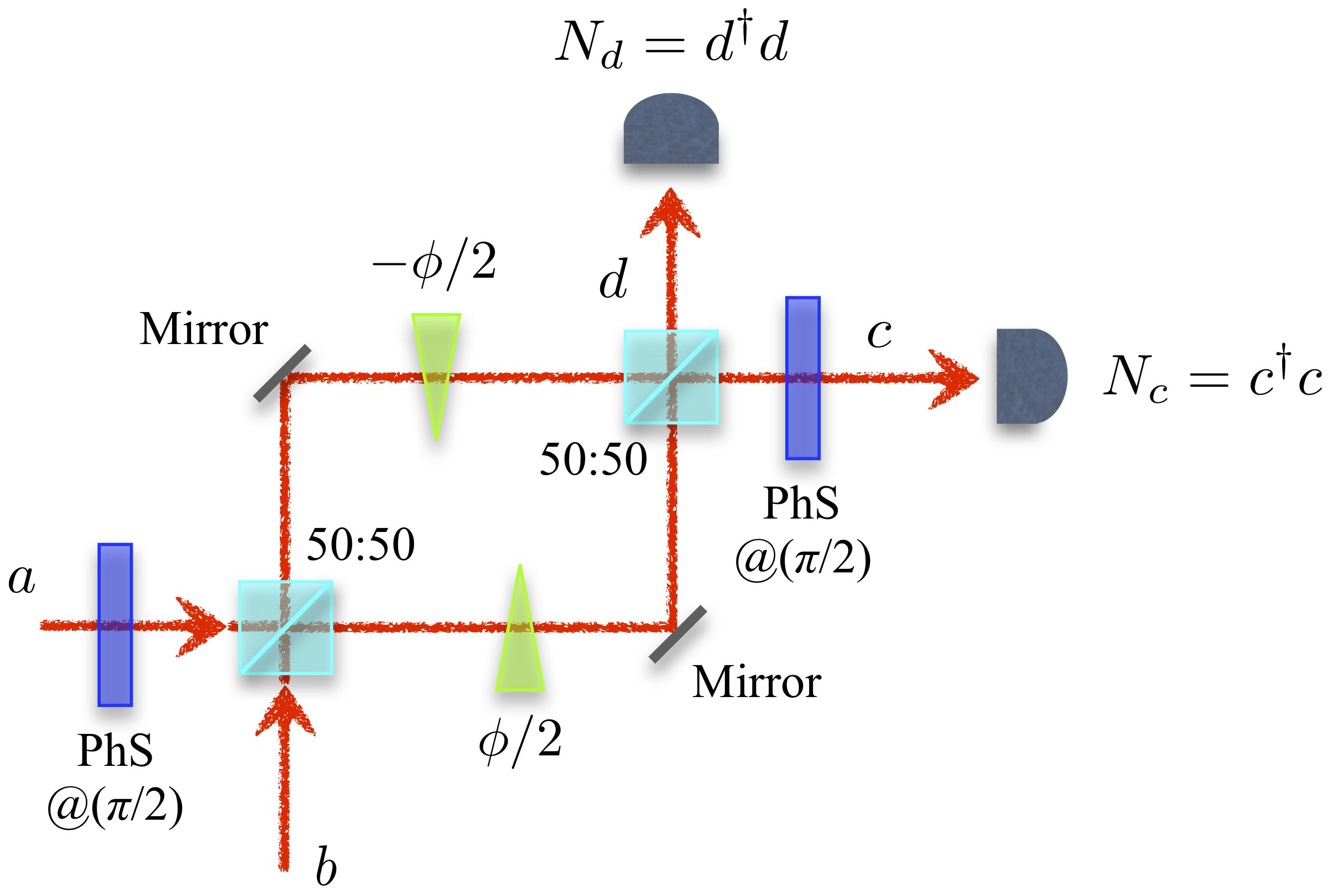}
\end{center}
\vspace{-0.3cm}
\caption{Scheme of a Mach-Zehnder interferometer. The two input modes,
described by the annihilation operators $a$ and $b$, with $[a,a^\dag] = {\mathbbm I}$
and $[b,b^\dag] = {\mathbbm I}$, interfere at a first 50:50 beam splitter and
an overall phase-shift difference $\phi$ between the beams is applied.
The two beams are then recombined at a second 50:50 beam splitter and, finally, they
are detected by measuring the number of photons $N_c = c^\dag c$ and
$N_d = d^\dag d$, respectively, where $c$ and $d$ are the annihilation
operators of the output fields, $[c,c^\dag] = {\mathbbm I}$ and $[d,d^\dag] = {\mathbbm I}$.
In the scheme we also added two phase shifters (PhS) at a fixed phase
$\pi/2$: this allows to describe the whole evolution of the input modes through the
interferometer as a simpler one in which they interfere at a single beam splitter with transmissivity
$\tau(\phi) = \cos^2(\phi/2)$, as shown in fig.~\ref{f:int:scheme}.}\label{f:MZ:scheme}
\end{figure}
The key ingredient of an optical interferometers is the beam splitter, a linear optical device
in which two beams of light interfere (see appendix~\ref{app:BS}).
In fig.~\ref{f:MZ:scheme} we sketched a Mach-Zehnder interferometer,
but analogous results can be obtained in the case of a Michelson interferometer.
The two input field modes, described by the annihilation operators
$a$ and $b$, with $[a,a^\dag] = {\mathbbm I}$
and $[b,b^\dag] = {\mathbbm I}$, evolve into the output modes $c$ and $d$
which depend on the phase $\phi$. To find the input-output relations,
we should transform the input modes through the first 50:50 beam splitter, then apply
the phase shift and the reflections at the mirrors and, finally, recombine the
resulting modes in a second 50:50 beam splitter.
\par
\begin{figure}[h!]
\begin{center}
\includegraphics[width=0.3\textwidth]{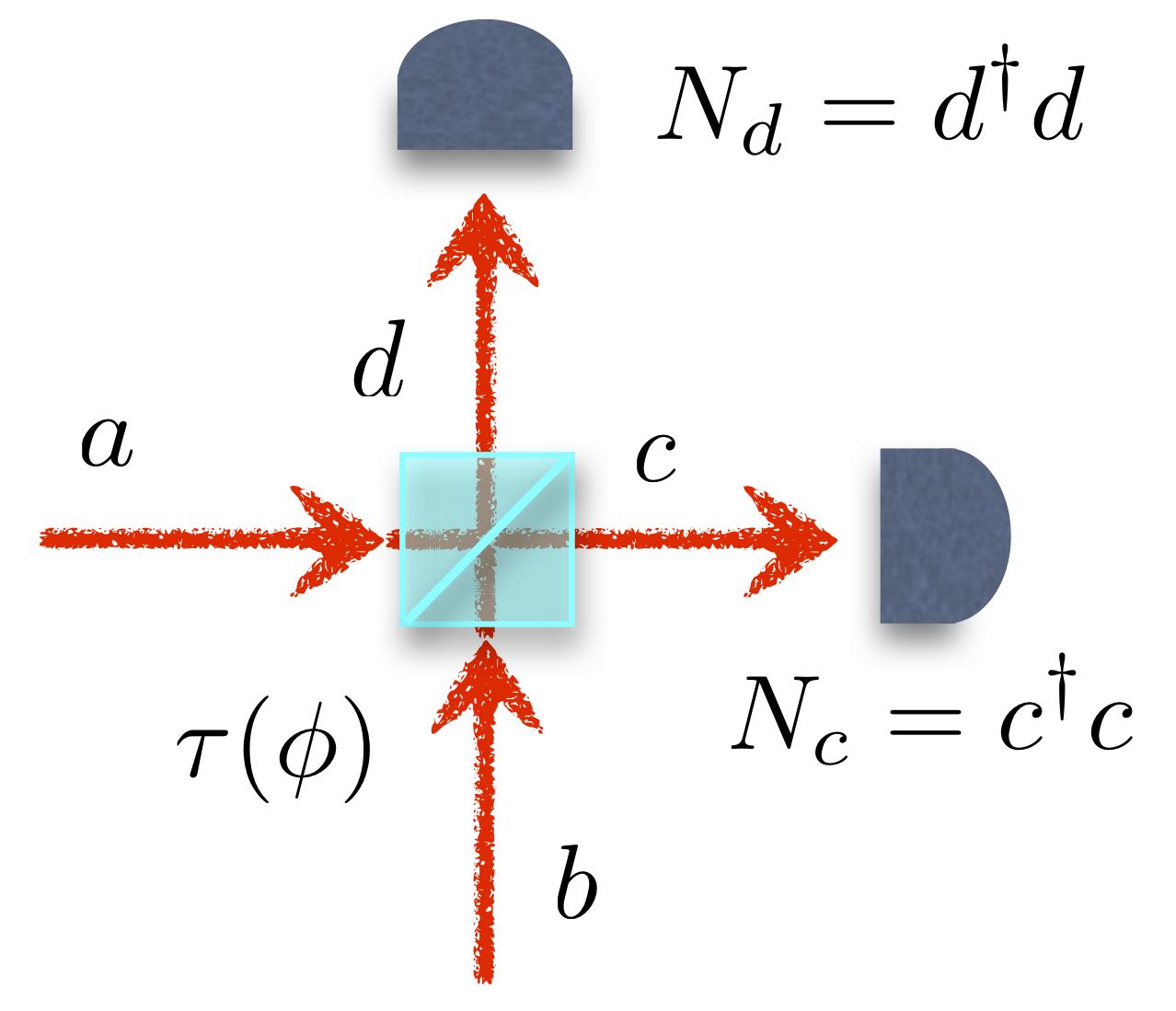}
\end{center}
\vspace{-0.3cm}
\caption{Simplified scheme of an interferometer equivalent to the
Mach-Zehnder configuration in fig.~\ref{f:MZ:scheme}.
The measured phase shift $\phi$ can be
summarised by the presence of a beam splitter transmissivity $\tau(\phi) = \cos^2(\phi/2)$.
The two input bosonic field modes, $a$ and $b$,
evolve into the output filed modes $c$ and $d$,
that are finally detected by measuring the number of photons $N_c$ and
$N_d$, respectively.}\label{f:int:scheme}
\end{figure}
Besides the final measurement, it is interesting to note that,
from the theoretical point of view, any interferometer measuring a phase shift $\phi$
can be summarised by two input light beams which interfere at a
beam splitter with transmissivity $\tau = \cos^2 (\phi/2)$, as sketched in fig.~\ref{f:int:scheme}.
In the case of the Mach-Zehnder interferometer of fig.~\ref{f:MZ:scheme},
this requires to add two phase shifters at $\pi/2$ (see PhS in fig.~\ref{f:MZ:scheme}).
In this view, as the reader can verify applying the transformations associated with the
beam splitter, with the mirror and with the phase shift described in
appendix~\ref{app:BS}, the input field modes evolve into the output modes:
\begin{eqnletter}
\label{in:out:int}
c &=& \cos(\phi/2)\, a + \sin(\phi/2)\, b\,,  \\[1ex]
d &=& \cos(\phi/2)\, b - \sin(\phi/2)\, a\,.
\end{eqnletter}
The final stage of an interferometer is the photodetection of the beams leading to the
measurement of the number of photons $N_c \equiv N_c (\phi)$ and $N_d \equiv N_d (\phi)$
and, finally, a suitable function $f(N_c, N_d)$ is evaluated (usually, as we will see later, the sum $N_c + N_d$
or the difference $N_c - N_d$ of the two quantities) obtaining a data sample.
The outcomes $x$ are distributed according to the probability distribution $p(x|\phi)$ which,
of course, depends on the states of the optical input beams.
The information about $\phi$ is retrieved using the estimator
\begin{equation}
{\cal O}(\phi) = \int_\Omega x\, p(x|\phi)\, dx\,,
\end{equation}
$\Omega$ being the data sample space, whereas the sensitivity $\sens$ of the interferometer is given by
the relation
\begin{equation}\label{snes:def}
\sens(\phi) = \frac{\sqrt{\var[{\cal O}(\phi)]}}{|\partial_\phi {\cal O}(\phi)|}.
\end{equation}
Whenever $p(x|\phi)$ can be approximated by a Gaussian distribution with standard deviation
$\sigma$, it easy to show that \cite{spara:PRA}
\begin{equation}
\sens(\phi) \approx \frac{1}{\sqrt{F(\phi)}} \approx \sigma\,,
\end{equation}
where
\begin{equation}
F(\phi) = \int_\Omega p(x|\phi)\, [ \partial_\phi \log p(x|\phi) ]^2\, dx
\end{equation}
is the Fisher information
introduced in section~\ref{s:QET}.
\par
In order to show how the presence of nonclassical states can improve the sensitivity,
we first consider the scheme in fig.~\ref{f:int:scheme} when a coherent state $| \alpha \rangle_a$ and the
vacuum state $| 0 \rangle_b$ enter the two ports of the interferometer.
We recall that the coherent state represents, with good approximation,
the output state of a laser with average number of photons given by $ N = |\alpha|^2$ and standard
deviation $|\alpha| = \sqrt{N}$ \cite{glauber} (see appendix~\ref{app:CS} for further details about coherent
states).
\par
\begin{figure}[h!]
\begin{center}
\includegraphics[width=0.95\textwidth]{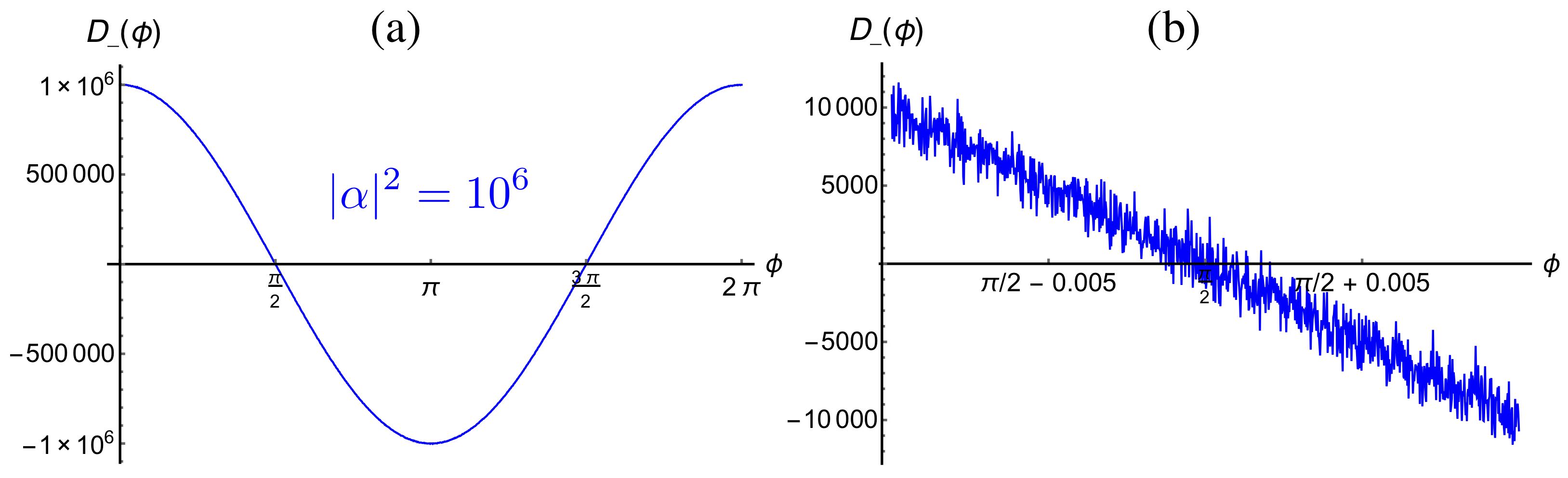}
\end{center}
\vspace{-0.3cm}
\caption{(a) Interference fringes (Monte Carlo simulated data).
Typical plot of the photon number difference $D_{-} (\phi)$ as a function
of $\phi$ at the output of an interferometer fed with a coherent state with amplitude $|\alpha|^2 = 10^6$
and the vacuum in the other input port. (b) Magnification of the region near $\phi = \pi/2$ of the
previous plot: we can see the noise due to the photon number fluctuations.}\label{f:int:fringes}
\end{figure}
If we assume to measure the photon number difference at the output, namely:
\begin{eqnletter}
{\cal O}(\phi) &=& D_{-}(\phi) \equiv \hbox{Tr}[\varrho_{{\rm in}} (N_c - N_d)]\,, \\[1ex]
&=& N_c(\phi) - N_d(\phi)\,,
\end{eqnletter}
with
\begin{equation}
\varrho_{{\rm in}} = | \alpha \rangle_a \langle \alpha | \otimes | 0 \rangle_b \langle 0 |\,,
\end{equation}
then we obtain the
typical interference fringes, as reported in fig.~\ref{f:int:fringes}~(a).
Since $D_{-}(\phi) = |\alpha|^2 \cos \phi$ and $\var[D_{-}(\phi)] = |\alpha|^2$,
it is easy to show that the sensitivity of this
interferometer is
\begin{equation}
\sens_{\rm cl}(\phi) = \frac{1}{|\alpha|}\frac{1}{|\sin\phi|}\,,
\end{equation}
(the subscript ``cl'' underlines the use of classical light,
such as the coherent state) which reaches the minimum for $\phi = \phi_0 = \pi/2$. 
The value of the phase leading to the minimum sensitivity is usually called ``working regime'' of the interferometer: here a small change of the phase produces a large effect on the quantity
$D_{-}(\phi)$, as we can see from fig.~\ref{f:int:fringes}~(b), since the absolute value of derivative at the denominator of eq.~(\ref{snes:def}) reaches its maximum.

\par
If we consider a scheme in which only one port of the interferometer is monitored, say the
port corresponding to mode $d$, one finds analogous results, but now
\begin{eqnletter}
{\cal O}(\phi) &=& N_d(\phi)\\[1ex]
&=& |\alpha|^2 \sin^2(\phi/2)\,,
\end{eqnletter}
and $\var[N_d(\phi)] = |\alpha|^2 \sin^2(\phi/2)$; therefore, the sensitivity becomes:
\begin{equation}
\sens_{\rm cl}(\phi) = \frac{1}{|\alpha|}\frac{1}{|\cos(\phi/2)|}\,.
\end{equation}
Note that in this case the working regime is $\phi = 0$, corresponding to
the so-called ``dark port'' as expected (we will briefly consider this configuration at the end of this
section). This is the typical working regime of gravitational wave antennas \cite{grav:11}. In both the cases, however, the minimum sensitivity scales as $\propto N^{-1/2}$, $N$
being the average number of photons circulating in the interferometer.
This is the well-known \emph{shot-noise scaling}. We will see that this limit can be beaten
exploiting nonclassical optical states up to reaching the \emph{Heisenberg scaling}
$\propto N^{-1}$ \cite{DDR:15,spara:josab,paris:sq,optimized:07,pezze:08,lang:13}.
\par
\begin{figure}[h!]
\begin{center}
\includegraphics[width=0.5\textwidth]{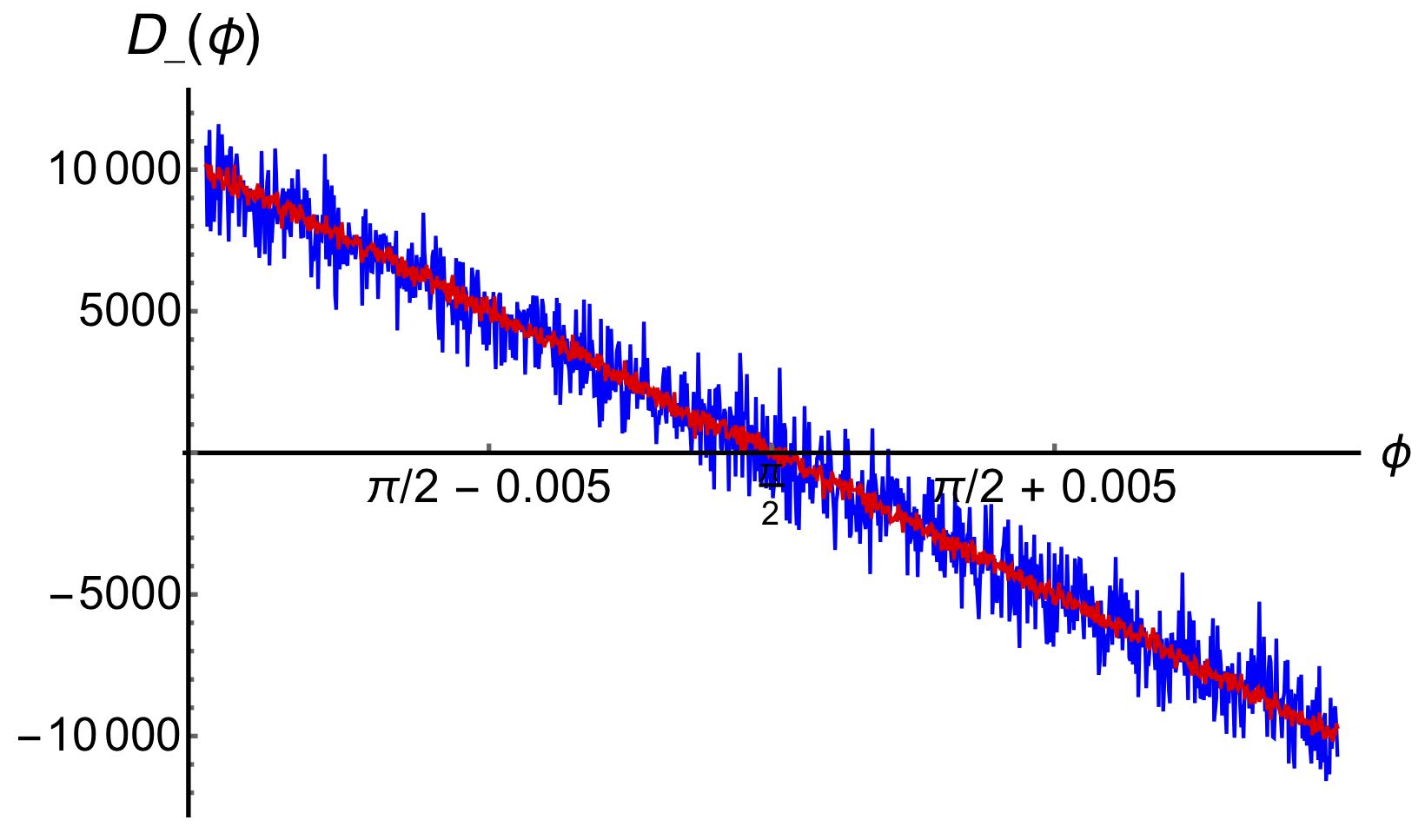}
\end{center}
\vspace{-0.3cm}
\caption{Same plot as in fig.~\ref{f:int:fringes}~(b) (blue line) and with a squeezed vacuum $| 0, r \rangle$
instead of the vacuum, with number of squeezing photons $\lambda = \sinh^2 r = 4$, corresponding
to about 12.5~dB,
and the same coherent amplitude $|\alpha|^2 = 10^6$ (red line). Note that $\lambda \ll |\alpha|^2$.
We can see that
the photon number fluctuations are clearly reduced by the presence of squeezing.}\label{f:int:squeezing}
\end{figure}
In order to improve the sensitivity we should look for a way to increase the value of the
denominator of eq.~(\ref{snes:def}) and/or to reduces the fluctuations of the measured
quantity at the numerator. In this last case we can exploit the
nonclassical properties of a particular class of states called squeezed states
(see appendix~\ref{app:SQ} for some details about the definition and the main properties of
squeezed states).
\par
\begin{figure}[h!]
\begin{center}
\includegraphics[width=0.5\textwidth]{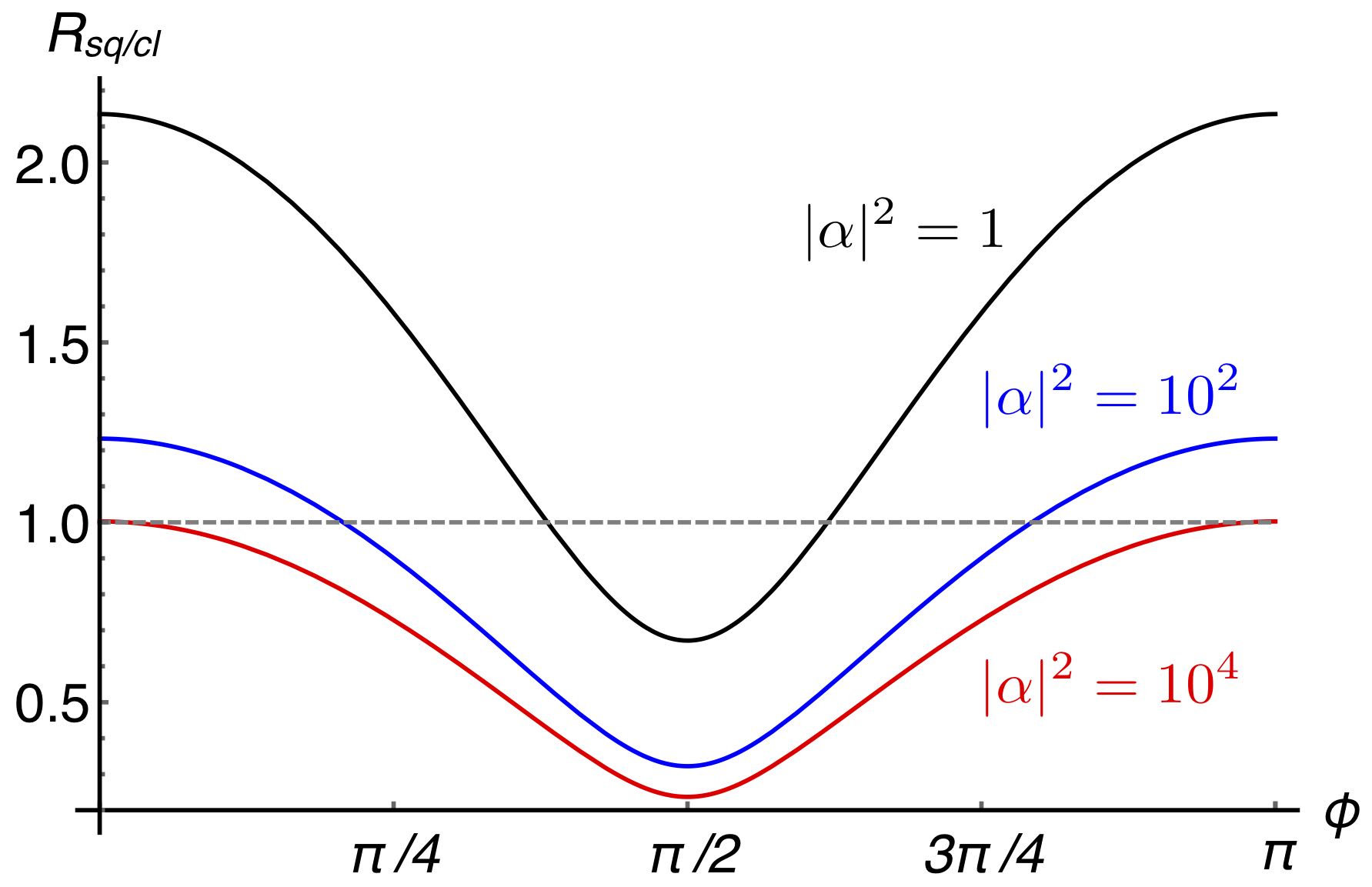}
\end{center}
\vspace{-0.3cm}
\caption{Plot of the ratio $R_{\rm sq/cl} = \sens_{\rm sq}(\phi)/\sens_{\rm cl}(\phi)$
between the sensitivity in the presence of squeezing
($\sens_{\rm sq}$) and without it ($\sens_{\rm cl}$) as a function of the measured phase $\phi$.
We used the same squeezing
parameter as in fig.~\ref{f:int:squeezing} and different values of $|\alpha|^2$. When $R_{\rm sq/cl} <1$
squeezing turns out to be a useful resource for phase estimation.}\label{f:int:ratio}
\end{figure}

If we substitute to the vacuum a \emph{squeezed} vacuum state, the two-mode input state reads
\begin{equation}
\varrho_{{\rm in}} = | \alpha \rangle_a \langle \alpha | \otimes | 0,r \rangle_b \langle 0,r |\,,
\end{equation}
where $| 0,r \rangle_b$ is a squeezed state with $0$ coherent amplitude and squeezing parameter $r$
(see appendix~\ref{app:SQ}).  The analytical formula of the sensitivity is now quite cumbersome \cite{spara:PRA} and is not reported explicitly. Nevertheless, fig.~\ref{f:int:squeezing} shows the effects
of the presence of squeezing, {\it i.e.} a noncalssical resource, on the interference fringes.
Though the number of squeezed photon for the chosen parameters is extremely small with
respect to number of coherent photons ($\lambda =  \sinh^2 r = 4 \ll |\alpha|^2 = 10^6$),
we can see a reduction of the photon
number fluctuation which improves the sensitivity of the interferometer \cite{paris:sq}.
This is clear from fig.~\ref{f:int:ratio}, where we plot the ratio
\begin{equation}
R_{\rm sq/cl} = \frac{\sens_{\rm sq}(\phi)}{\sens_{\rm cl}(\phi)}
\end{equation}
between the sensitivity in the presence of squeezing ($\sens_{\rm sq}$) and without it
($\sens_{\rm cl}$) as a function of the measured phase $\phi$.
It is worth noting that only for $\phi \approx \pi/2$ the use of squeezing leads to
improved interferometer performances.
This can be understood by inspecting the explicit expression of the variance
of $D_{-}(\phi)$, which is now phase sensitive because of the presence of squeezing and reads
\begin{equation}
\var[D_{-}(\phi)] = |\alpha|^2 
\left[
1 - 2 \sqrt{\lambda}\left(\sqrt{\lambda+1} -\sqrt{\lambda} \right) \sin^2 \phi
\right]
+\lambda \left[
1 + (1+2\lambda) \cos^2 \phi
\right]\,.
\end{equation}
Therefore, we have:
\begin{equation}
|\alpha|^2 \left[
1 - 2 \sqrt{\lambda}\left(\sqrt{\lambda+1} -\sqrt{\lambda} \right) \right] + \lambda
 \le
\var[D_{-}(\phi)] 
\le
|\alpha|^2 + 2 \lambda (1 + \lambda)\,,
\end{equation}
where the minimum is achieved for $\phi = \pi/2$ and the maximum for $\phi = 0, \pi$.
For what concerns the expectation value of the photon number difference, the
squeezing just affects the amplitude of the interference fringes, namely:
\begin{equation}
D_{-}(\phi) = \left( |\alpha|^2 - \lambda \right) \cos \phi\,.
\end{equation}
Note that, in order to have $\var[D_{-}(\pi/2)] <  |\alpha|^2$, we should require
$\lambda < 4 |\alpha|^2 / (1 + 4 |\alpha|^2)$, otherwise the squeezing resource becomes useless.
Moreover, in the regime $\lambda \ll |\alpha|^2$ (see the lower red line in fig.~\ref{f:int:ratio})
we find:
\begin{equation}
\var[D_{-}(\phi)] \approx |\alpha|^2 
\left[
1 - 2 \sqrt{\lambda}\left(\sqrt{\lambda+1} -\sqrt{\lambda} \right) \sin^2 \phi
\right]\,,
\end{equation}
and we can see that it always beats the performance of an interferometer fed by the
a coherent state (and, of course, the vacuum).
\begin{figure}[h!]
\begin{center}
\includegraphics[width=0.95\textwidth]{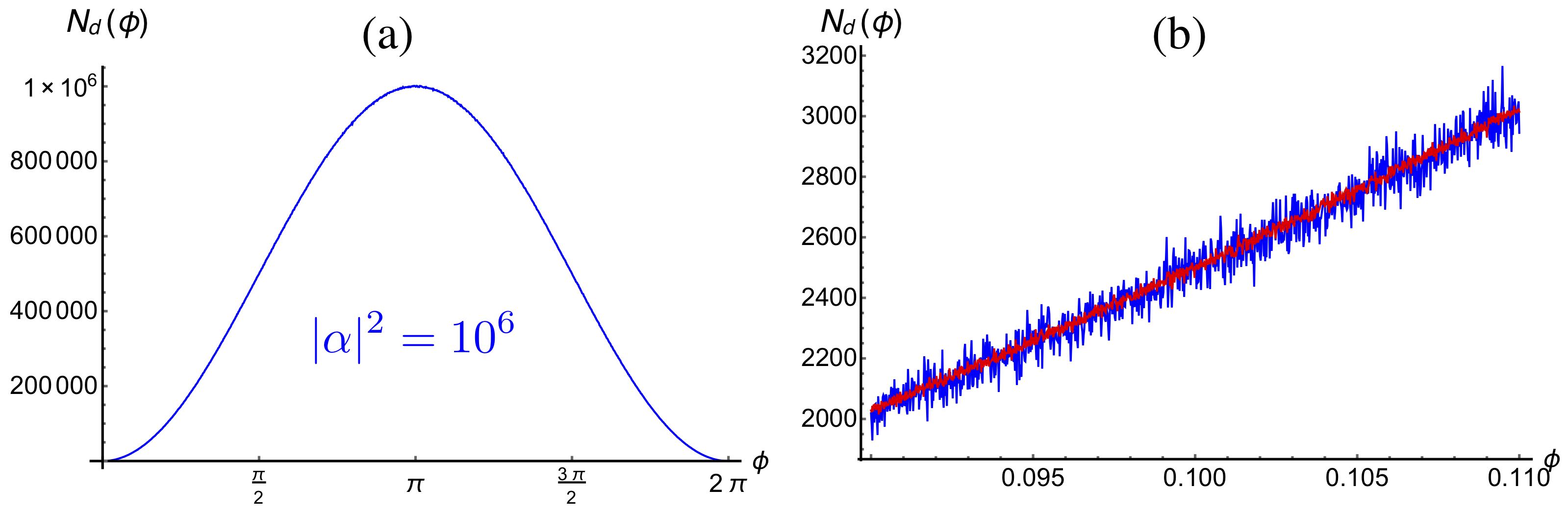}
\end{center}
\vspace{-0.3cm}
\caption{(a) Interference fringes (Monte Carlo simulated data) when only the port corresponding
to mode $d$ (see fig.~\ref{f:int:scheme}) is monitored.
Typical plot of the photon number $N_d (\phi)$ as a function
of $\phi$ at the output of an interferometer fed with a coherent state with amplitude
$|\alpha|^2 = 10^6$ and the vacuum in the other input port. (b) Magnification of the region
approaching $\phi = 0$: we can see the noise due to the photon number fluctuations in the
presence of coherent state and vacuum (blue line) and their reduction when the vacuum state
is substituted with a squeezed vacuum (red lines).}\label{f:int:fringes:dark}
\end{figure}
\begin{figure}[h!]
\begin{center}
\includegraphics[width=0.5\textwidth]{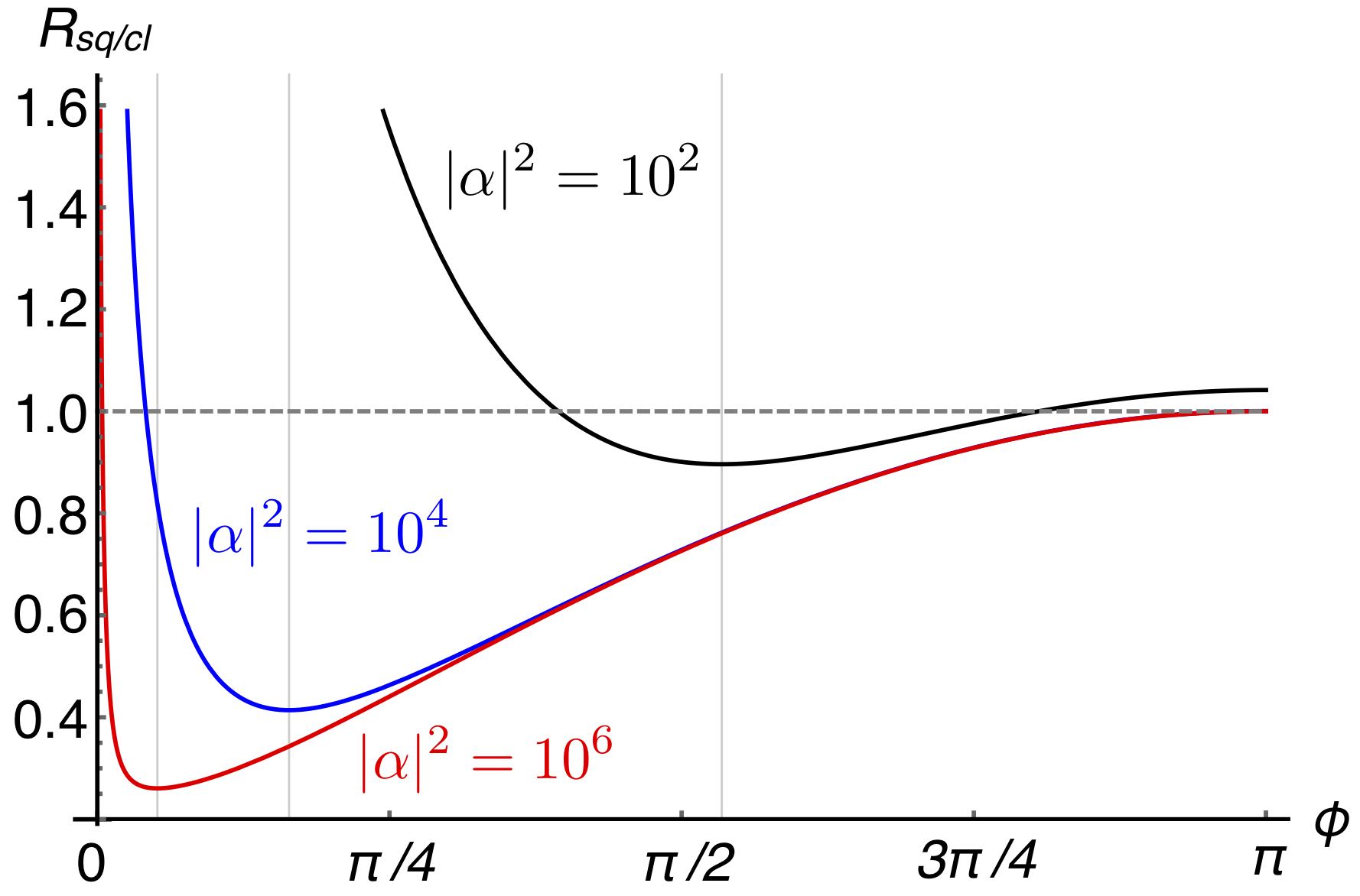}
\end{center}
\vspace{-0.3cm}
\caption{Plot of the ratio $R_{\rm sq/cl} = \sens_{\rm sq}(\phi)/\sens_{\rm cl}(\phi)$
between the sensitivity in the presence of squeezing
($\sens_{\rm sq}$) and without it ($\sens_{\rm cl}$) as a function of the measured phase $\phi$
when only the port of mode $d$ (see fig.~\ref{f:int:scheme}) is monitored.
We used the same squeezing
parameter as in fig.~\ref{f:int:squeezing} and different values of $|\alpha|^2$. When $R_{\rm sq/cl} <1$
squeezing turns out to be a useful resource for phase estimation. Note the presence of the minimum
approaching $\phi = 0$ as the coherent state amplitude increases.}\label{f:int:ratio:dark}
\end{figure}
For the sake of completeness in fig.~\ref{f:int:fringes:dark}~(a) we report the interference fringes
when only one port of the interferometer is monitored  and the number of photons $N_d(\phi)$
is recorded. In this case, as we noted above, the working regime is
$\phi = 0$, where the intensity fluctuations, which depends on the coherent state amplitude
(see appendix~\ref{app:CS}), vanishes, since the detected intensity is null.
In fig.~\ref{f:int:fringes:dark}~(b), blue line, we show the magnification of the region
for $\phi$ approaching zero. As in the case of the measurement of $D_{-}(\phi)$,
also in this one the presence of squeezing can help: the red line in fig.~\ref{f:int:fringes:dark}~(b)
is obtained by substituting to the vacuum a squeezed state with the same squeezing parameter
used in fig.~\ref{f:int:squeezing}.
\par
If we now consider the corresponding ratio $R_{\rm sq/cl}$
between the sensitivity in the presence of squeezing and without it reported in fig.~\ref{f:int:ratio:dark}, we can
see that the minimum is not achieved in the working regime $\phi = 0$, but for a value that approaches
it as the intensity of the coherent state increases.
On the contrary, we recall that when the photocurrent difference
$D_{-}(\phi)$ at the output is used to retrieve the information about the phase, the working regime
at $\phi = \pi/2$ coincides with the optimal phase to fully exploit the squeezing resource,
as we can see from fig.~\ref{f:int:squeezing} and fig.~\ref{f:int:ratio}.
\par
Also in this case there are regions of the phase for which squeezing does not
improve the performance of the interferometer. However, the analysis of the effect is more
subtile than the previous case. In fact, while the mean value reads
\begin{equation}
N_d (\phi) = |\alpha|^2 \sin^2(2\phi) + \lambda \cos^2(2\phi)\,,
\end{equation}
the variance is now given by
 \begin{eqnarray}
\var[N_{d}(\phi)] &=& \frac{|\alpha|^2}{2}
\left[
1 - \cos\phi +  \sqrt{\lambda}\left(\sqrt{\lambda+1} -\sqrt{\lambda} \right) \sin^2 \phi
\right]\\
 & & \hspace{3cm}+  \frac{\lambda}{2}\cos^2\left(\frac{\phi}{2}\right)\left[
2 + (1 + 2\lambda)(1+\cos\phi)
\right]\,, \notag
\end{eqnarray}
which depends on $\phi$ also in the absence of squeezing. If we focus on the working
regime $\phi=0$, we have $\var[N_{d}(0)] = 2 \lambda (1+\lambda)$, which vanishes only
for $\lambda =0$, {\it i.e.} when the squeezed state is replaced by the vacuum. Nevertheless,
squeezing can help in realistic scenarios, in which one cannot set precisely the phase
$\phi=0$ because of the unavoidable experimental errors.
As shown in fig.~\ref{f:int:ratio:dark} (see the red curve for
$|\alpha|^2 = 10^6$) it is enough a small deviation from $\phi=0$ to make the squeezing
a useful resource. In fact, when a very large number of photons
circulates in the interferometer, a phase shift slightly different from $0$
is enough to bring  a considerable number of photons to the detector. In this case the presence of
squeezing can reduce the intensity fluctuation and, in turn, increase the sensitivity.
\par
In the next section, we will investigate the limits imposed by quantum mechanics to
the precision of optical interferometers by using the tools of quantum estimation theory
introduced in section~\ref{s:QET}. Moreover, we will also consider more general setups involving
active devices, such as optical parametric amplifiers.

\section{Bounds to precision for quantum interferometry}\label{s:bounds}

\begin{figure}[h!]
\begin{center}
\includegraphics[width=0.6\textwidth]{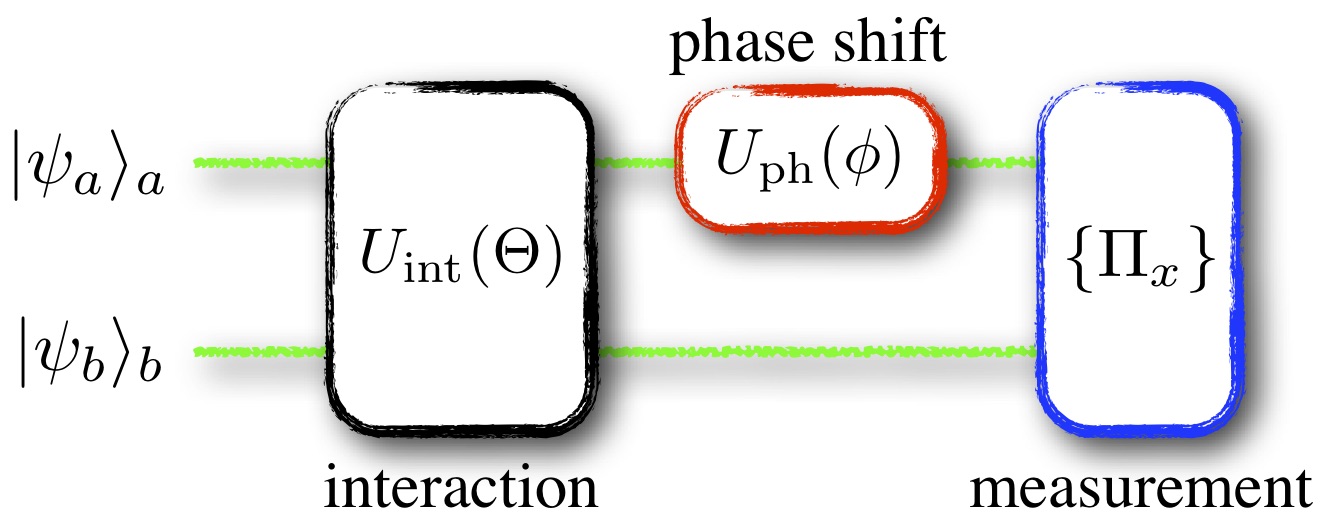}
\end{center}
\vspace{-0.3cm}
\caption{Diagram of an interferometer showing the main elements: the unitary interaction $U_{\rm int}(\Theta)$,
the phase shift $U_{\rm ph}(\phi)$ and the measurement described by the POVM $\{ \Pi_x \}$.}\label{f:int:param:scheme}
\end{figure}

When a particular measurement stage is chosen and the input states are given,
in order to find the optimal working regime
of an interferometer we should maximise the Fisher information with respect
to all the involved parameters (characterising both the input states and the interferometer).
Moreover, we should address and optimise the quantum Fisher information
\cite{QCR:94,QCR:96} to study the ultimate bounds imposed by quantum mechanics,
and find the optimal measurement to perform.
To this aim it is useful to describe the interferometers as sketched in fig.~\ref{f:int:param:scheme}.
Here, the two modes of an input state
\begin{equation}
| \Psi \rangle\rangle = | \psi_a \rangle_a | \psi_b \rangle_b
\end{equation}
interacts through
the unitary interaction $U_{\rm int}(\Theta)$, $\Theta$ representing the parameter of the interaction
(such as the beam splitter transmissivity), then one of the modes undergoes a phase shift
$U_{\rm ph}(\phi)$; finally,
they are measured according to the POVM $\{ \Pi_x \}$, $x$ being the outcome of the measurement \cite{paris:rev}.
\par
In general, different quantum optical states can be used to improve the sensitivity of
an interferometer \cite{QM,kok:int,popo}. In this section we consider the class Gaussian states,
namely, states described by Gaussian Wigner function
which can be generated and manipulated with the current technology \cite{bachor,porzio:09,cialdi:16}.
Therefore we can assume that the input states are two displaced squeezed
states, {\it i.e.} $| \alpha, r  \rangle | \gamma , \xi \rangle$
(for the sake of simplicity we drop the subscripts), where
we can also assume $\alpha, \gamma, r, \xi \in {\mathbbm R}$ \cite{spara:PRA}. In order to make the analysis
easier, it is useful to introduce the following relevant quantities:
\begin{eqnletter}
N_{\rm tot} = \alpha^2 + \gamma^2 +\sinh^2 r + \sinh^2 \xi\,,&
&(\hbox{total number of photons}) \nonumber \\
\Delta = {\displaystyle \frac{\alpha^2}{ \alpha^2+ \gamma^2}\,,}&
&(\hbox{signal fraction}) \nonumber \\
\beta_{\rm tot} = {\displaystyle \frac{\sinh^2 r + \sinh^2 \xi}{N_{\rm tot}}\,,}&
&(\hbox{total squeezing fraction}) \nonumber \\
\beta = {\displaystyle \frac{\sinh^2 \xi}{N_{\rm tot}}\,.}&
&(\hbox{total squeezing fraction}) \nonumber 
\end{eqnletter}
As we discussed in section~\ref{s:QET},
the inverse of the Fisher information sets a lower bound for the the variance in estimating $\phi$,
known as Cram\'er-Rao bound, that is based on $p(x|\phi)$ and, thus,
depends on the particular measurement we perform. It is possible, however, to find a measurable
observable (an optimal POVM) that maximises the Fisher information \cite{paris:rev,helstrom,QFI}.
This leads to an upper bound for the Fisher information, the \emph{quantum Fisher information}
(see section~\ref{s:QET}).
\par
In order to find the optimal POVM one starts from the Born rule,
\begin{equation}
p(x|\phi) = \hbox{Tr}[\varrho_{\phi} \Pi_x]\,,
\end{equation}
where
\begin{equation}
\varrho_{\phi} =
U_{\rm ph}(\phi) U_{\rm int}(\Theta) \varrho_{\rm in} U_{\rm int}^{\dag}(\Theta) U_{\rm ph}^{\dag}(\phi)\,,
\end{equation}
and $\varrho_{\rm in} =  | \Psi \rangle\rangle \langle\langle \Psi |$.
To find the optimal $\{\Pi_x\}$ we write the Fisher information as in eq.~(\ref{Fisher:SLD}) and
the corresponding quantum Fisher information $H(\phi)$ is such that $H(\phi) \ge F(\phi)$.
Here we are not interested in the actual form of the POVM, but we will focus on the
quantum Fisher information.
\par
\begin{figure}[h!]
\begin{center}
\includegraphics[width=0.95\textwidth]{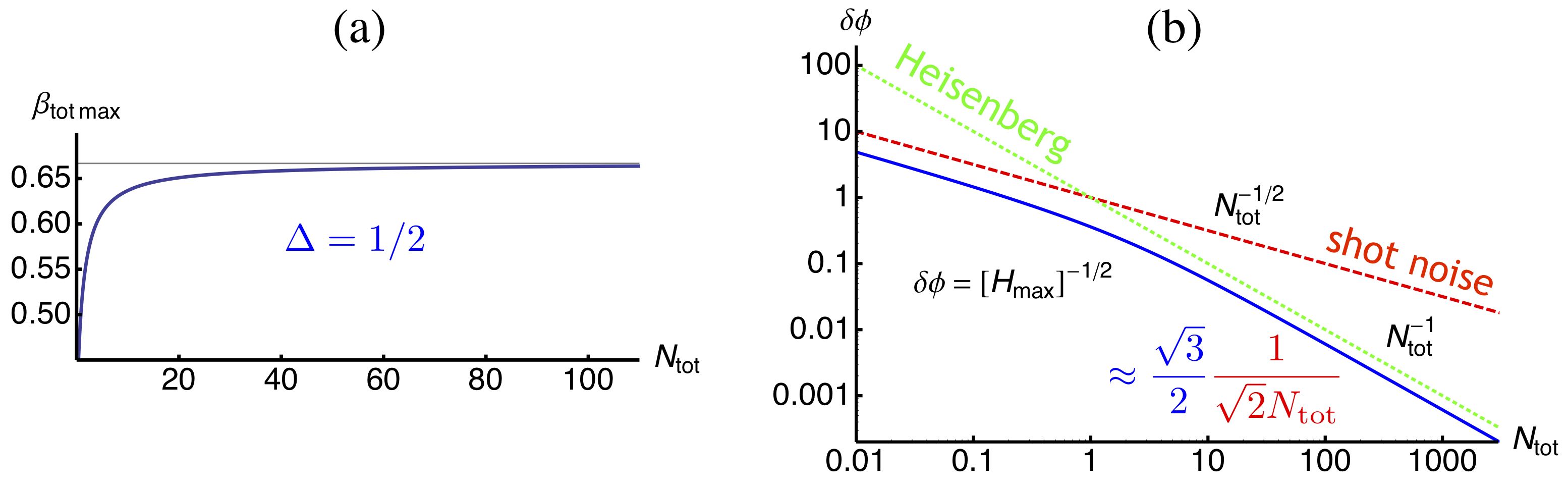}
\end{center}
\vspace{-0.3cm}
\caption{(a) Plot of the optimal value of $\beta_{\rm tot}$ maximising the
quantum Fisher information (\ref{QFI:passive}) as a function of $N_{\rm tot}$.
As $N_{\rm tot}$ increases $\beta_{\rm tot max} \to 2/3$.
(b) The phase sensitivity $\delta \phi = [H_{\rm max}]^{-1/2}$
as a function of $N_{\rm tot}$ (blue solid line), 
where we used the optimal value $\beta_{\rm tot\, max}$ given in panel (a).
In the limit $N_{\rm tot} \gg 1$
we have $H_{\rm max} \approx \frac{8}{3} N_{\rm tot}^2$, that is the Heisenberg
scaling. We also plot the shot-noise limit (red dashed line) and the Heisenberg limit
(green dotted line).}\label{f:QFI:passive:opt}
\end{figure}
First of all we consider a passive interaction, namely, an interaction which does not change the
energy of the input states. In particular, we assume that $U_{\rm int} (\Theta)$ describes the action of
a balanced beam splitter (see appendix~\ref{app:BS}):
\begin{equation}
U_{\rm int} (\Theta) \equiv \exp\left[\frac{\pi}{4}(a^\dag b -a b^\dag)\right]\,,
\end{equation}
where $a$ and $b$ are the boson field operators describing the two input modes. The maximisation of the
Fisher information over all the involved quantities introduced above shows that the best configuration
requires $\alpha = \gamma$ and $\xi = r$, corresponding to the following quantum Fisher information
(the reader can find the details in ref.~\cite{spara:josab}),
which depends only on $N_{\rm tot} = 2 (\alpha^2  + \sinh^2 r)$ and $\beta_{\rm tot} = 2 \sinh^2 r / N_{\rm tot}$:
\begin{eqnarray}
\label{QFI:passive}
H_{\rm max} \!\!&(&\!\! N_{\rm tot},\beta_{\rm tot}) = \\
&2&\!\! N_{\rm tot} \left[ 2 N_{\rm tot} \beta_{\rm tot}(2-\beta_{\rm tot}) +
2(1-\beta_{\rm tot})\sqrt{N_{\rm tot}\beta_{\rm tot}(2+N_{\rm tot}\beta_{\rm tot})}
\right]\,. \nonumber
\end{eqnarray}
We can perform a further (numerical) maximisation of $H_{\rm max} (N_{\rm tot},\beta_{\rm tot})$
with respect to the total squeezing fraction $\beta_{\rm tot}$.
The results are shown in fig.~\ref{f:QFI:passive:opt}~(a),
where we plot $\beta_{\rm tot\, max}$ as a function of $N_{\rm tot}$,
whereas in fig.~\ref{f:QFI:passive:opt}~(b) we plot the corresponding phase sensitivity
$\delta\phi = 1/\sqrt{H_{\rm max}}$
obtained by using the optimal squeezing fraction $\beta_{\rm tot\, max}$:
it is clear that in the large energy regime ($N_{\rm tot} \gg 1$) the Heisenberg scaling is reached.
\par
It is worth noting that the maximisation over all the parameters leads to the optimal
symmetric input state  $| \alpha, r  \rangle | \alpha , r \rangle$. If we assume, however,
that the input state has the form  $| \alpha, 0  \rangle | 0, \xi \rangle$ (coherent state $+$
squeezed vacuum \cite{grav:11}), the optimisation still gives the Heisenberg scaling for
$N_{\rm tot} \gg 1$ but now $H_{\rm max} \approx 2 N_{\rm tot}$, that is a slightly worse
result than the one we obtained by considering displaced squeezed states.
\par
Let's suppose, now, that the initial interaction is an active one, namely it increases the
energy of the input states. A typical interaction of this kind is implemented by
an optical parametric amplifier (OPA) \cite{oli:rev} and it is the so-called two-mode
squeezing interaction (see appendix~\ref{app:TWB}), described by the unitary operator:
\begin{equation}
U_{\rm int} (\Theta) \equiv \exp(\xi a^\dag b^\dag -\xi^*a b)\,.
\end{equation}
This operator, when
applied to the vacuum state, generates the two-mode squeezed vacuum or
twin-beam state, as explicitly shown in appendix~\ref{app:TWB}, namely:
\begin{equation}
| \hbox{TWB} \rangle\rangle = \sqrt{1-|\lambda|^2} \sum_n \lambda^n | n \rangle_a | n \rangle_b\,,
\end{equation}
with $\lambda = e^{\arg[\xi]} \tanh |\xi|$,
that is a continuous-variable entangled state whose nonclassical features are
exploited in many continuous-variable quantum information processing protocols
\cite{rev:gauss}.
\par
\begin{figure}[h!]
\begin{center}
\includegraphics[width=0.95\textwidth]{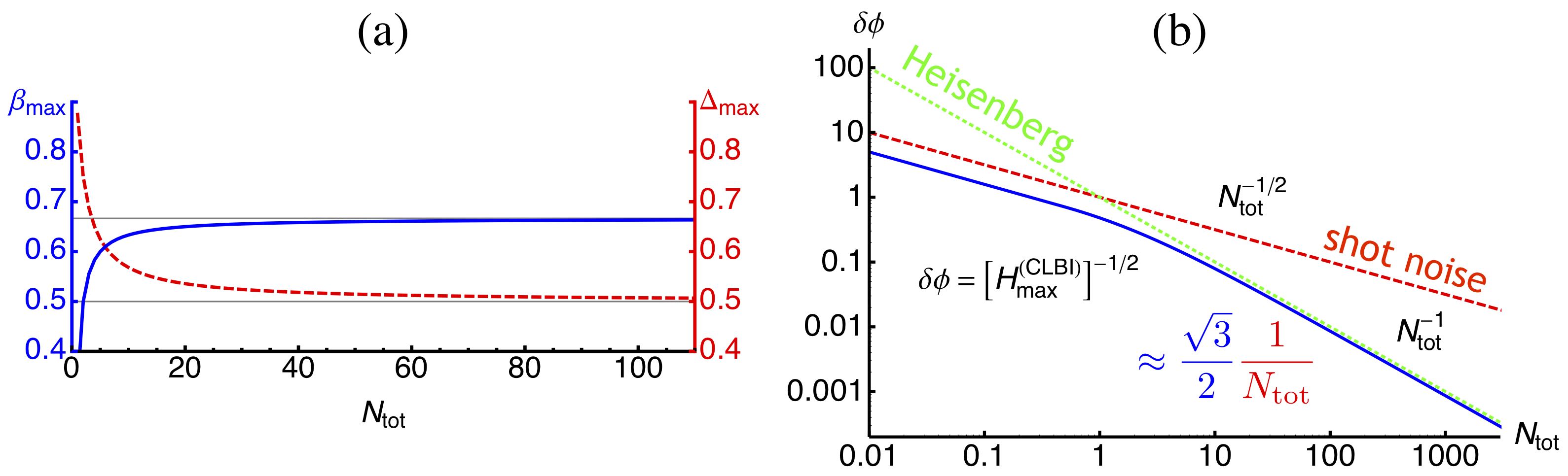}
\end{center}
\vspace{-0.3cm}
\caption{(a) Plot of $\beta_{\rm max}$ (blue sold line)
and $\Delta_{\rm max}$ (red dashed line) maximising the
quantum Fisher information in the case of the CLBI as function
$N_{\rm tot}$. In the limit $N_{\rm tot} \gg 1$ we have $\beta_{\rm max} \to 2/3$
and $\Delta_{\rm max} \to 1/2$.
(b) The corresponding phase sensitivity $\delta \phi = [H_{\rm max}^{\rm (CLBI)}]^{-1/2}$
as a function of $N_{\rm tot}$ (blue solid line). In the limit $N_{\rm tot} \gg 1$
we have $H_{\rm max} \approx \frac{4}{3} N_{\rm tot}^2$, that is the Heisenberg
scaling. We also plot the shot-noise limit (red dashed line) and the Heisenberg limit
(green dotted line).}\label{f:QFI:active:opt}
\end{figure}
This kind of interferometer is also referred to as \emph{coherent light boosted
interferometer} (CLBI) \cite{yurke:86,plick:10}. Since the interaction imposes
phase-sensitive amplification ({\it i.e.} two-mode squeezing) introducing 
quantum correlations between the two modes, we can choose as inputs two classical signals,
namely, a couple of coherent states
$| \alpha \rangle_a | \gamma \rangle_b$, $\alpha, \gamma \in {\mathbbm R}$.
Now the relevant quantities are:
\begin{eqnletter}
N_{\rm tot} = (\alpha^2 + \gamma^2 +1) \cosh(2r) +2\alpha\gamma \cos(\theta)\sinh(2 r)-1\,,&
&(\hbox{total number of photons}) \nonumber \\
\Delta = \frac{\alpha^2}{ \alpha^2+ \gamma^2}\,,&
&(\hbox{signal fraction}) \nonumber \\
\beta = \frac{\sinh^2 \xi}{N_{\rm tot}}\,. &
&(\hbox{total squeezing fraction}) \nonumber 
\end{eqnletter}
Proceeding as in the case of the passive interferometer, we can maximise the
quantum Fisher information with respect to $\beta$ and $\Delta$ for fixed energy
$N_{\rm tot}$. In fig.~\ref{f:QFI:active:opt}~(a) we plot
the signal fraction $\Delta_{\rm max}$ and the squeezing fraction $\beta_{\rm max}$
giving the maximal quantum Fisher information $H_{\rm max}^{\rm (CLBI)}$
plotted in fig.~\ref{f:QFI:active:opt}~(b). By comparison between fig.~\ref{f:QFI:passive:opt}~(b)
and fig.~\ref{f:QFI:active:opt}~(b) in the large energy regime we can see that
the passive interferometer outperforms the active one. Nevertheless,
in the presence of losses and realistic measurement the latter turns out to be useful,
as we are going to show in the following.
\par
A thorough analysis of passive and active interferometers using passive and active
detection stage can be found in ref.~\cite{spara:PRA}, while here we summarise the
main results we obtained also in the presence of non-unit quantum detection efficiency
(in appendix~\ref{app:eta} we describe the model used to include the quantum efficiency
in our analysis).
As a matter of fact, one can choose any possible combination of active/passive
interaction and active/passive detection.
\par
\begin{figure}[h!]
\begin{center}
\includegraphics[width=0.8\textwidth]{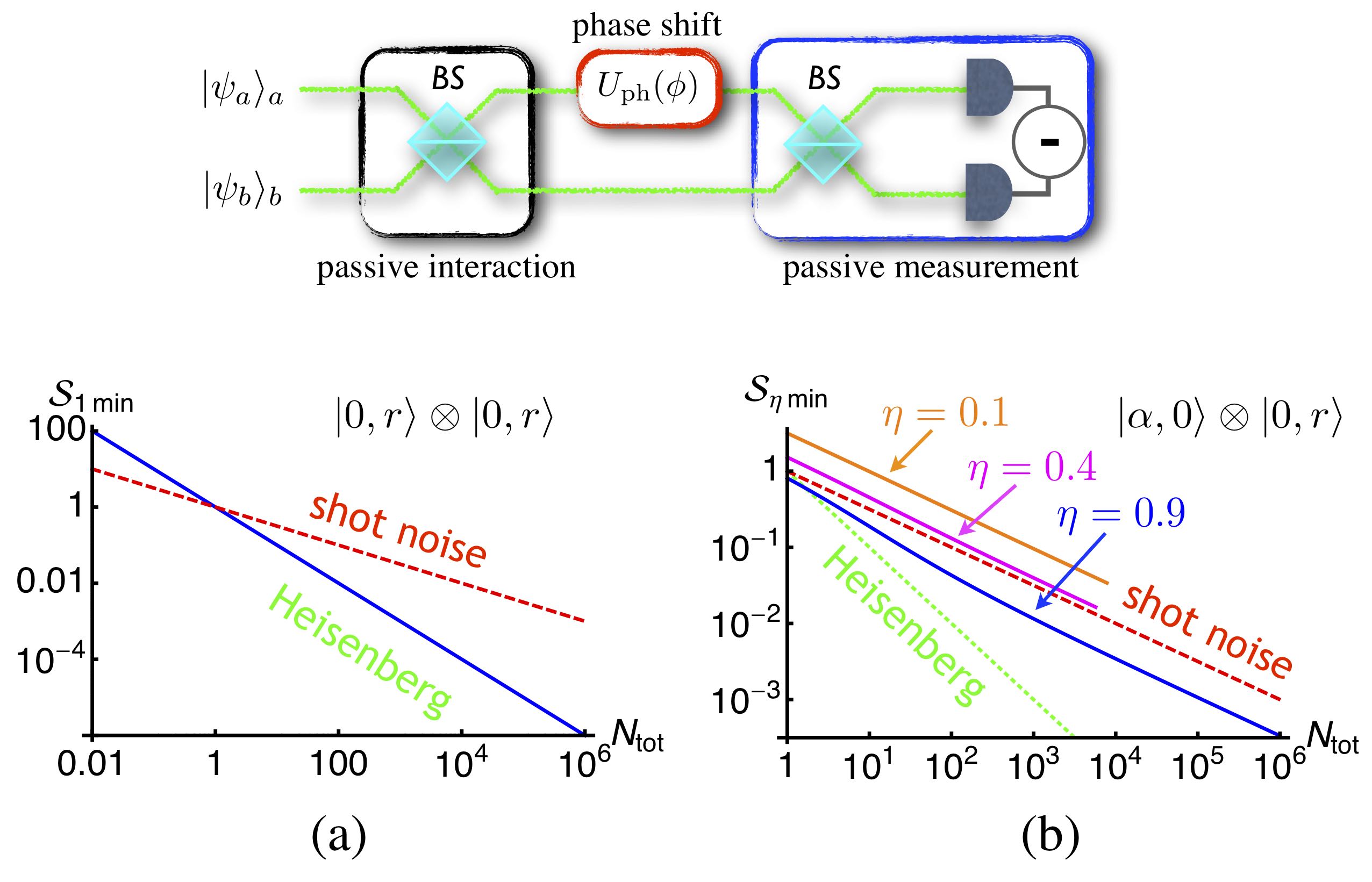}
\end{center}
\vspace{-0.5cm}
\caption{Interferometer with passive interaction and passive detection scheme: the two input
states are mixed at a balanced beam splitter (BS) before the phase shift and
interfere again at another BS before the photodetection. Here one evaluate the difference
$D_{-}(\phi) =N_1(\phi) - N_2(\phi)$ between the detected photons in order to retrieve
the information about $\phi$. (a) Minimised sensitivity of the interferometer for unit
quantum efficiency $\eta$ as a function of $N_{\rm tot}$. The optimal input states are
two squeezed vacua. (b) Minimised sensitivity of
the interferometer for different values $\eta$. Now the optimal input states are
a coherent state and a squeezed vacuum (see ref.~\cite{spara:PRA} for further details).
The shot-noise and the Heisenberg scalings are also reported for comparisons.}
\label{f:passive:passive}
\end{figure}
We first consider the most common configuration, in which a balanced, 50:50, beam splitter is
used in the initial interaction and before photodetection, as shown in
the scheme in fig.~\ref{f:passive:passive}: the measured quantity is the
difference $D_{-}(\phi)$ and the sensitivity is given by
\begin{equation}
\sens_{1}(\phi) = \frac{\sqrt{\hbox{var}[D_{-}(\phi)]}}{|\partial_\phi D_{-}(\phi)|}\,,
\end{equation}
In fig.~\ref{f:passive:passive}~(a) we plot the sensitivity of the interferometer as a function of $N_{\rm tot}$
after the optimisation ({\it i.e.} the minimisation of the sensitivity) with respect to the input states.
In the presence of unit quantum efficiency, we find that the optimal input states are
a couple of squeezed vacuum states: this choice allows us to obtain the Heisenberg
scaling. However, as one may expect, when $\eta < 1$ we can only achieve the shot-noise
limit \cite{DDR:13} and, in this case, the optimal input states are a coherent state and the squeezed
vacuum (see ref.~\cite{spara:PRA} for further details) as reported in fig.~\ref{f:passive:passive}~(b).
\par
\begin{figure}[h!]
\begin{center}
\includegraphics[width=0.95\textwidth]{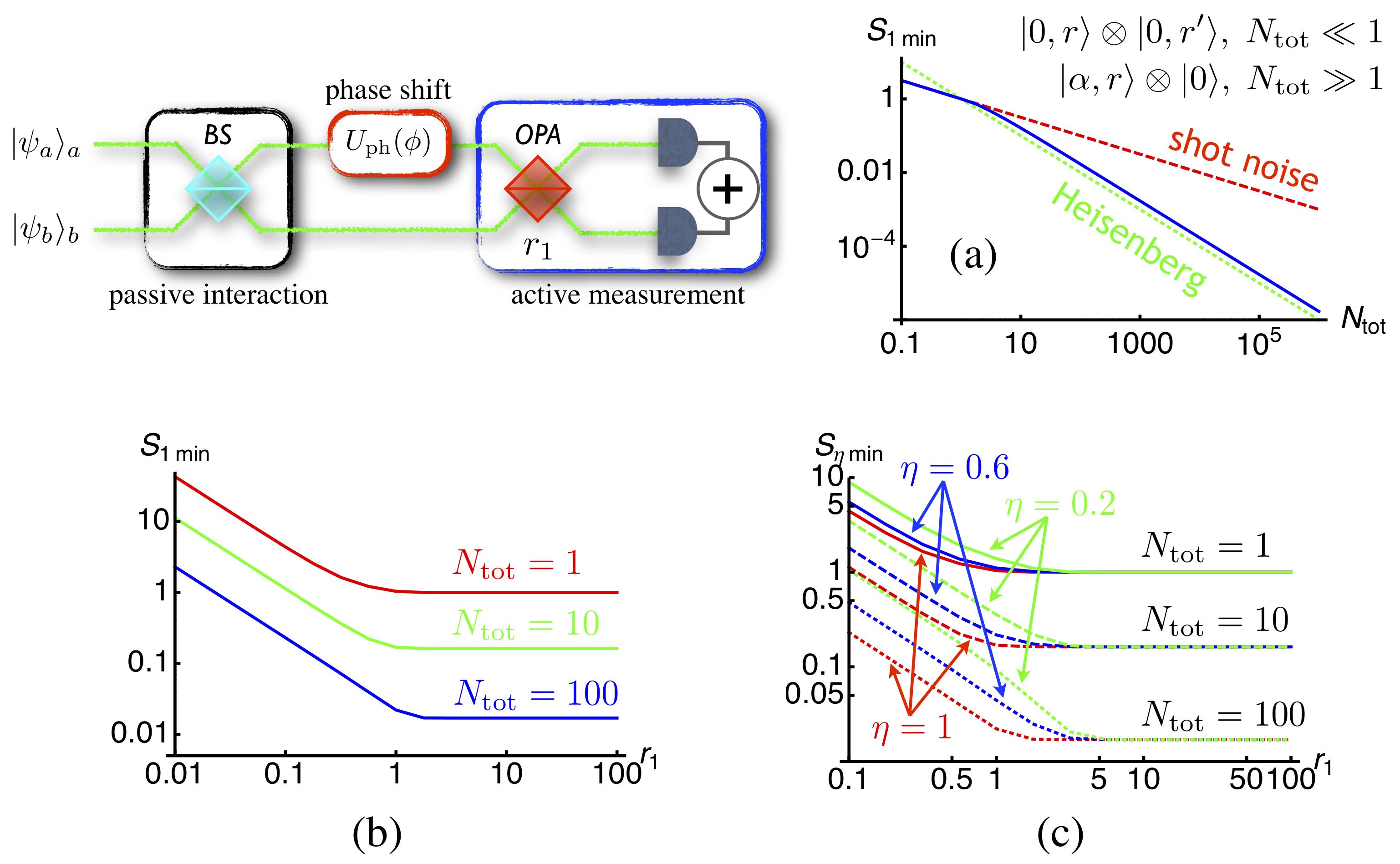}
\end{center}
\vspace{-0.5cm}
\caption{Interferometer with passive interaction and active detection stage:
the two input states are mixed at a balanced beam splitter (BS) before the phase
shift and, then, they undergo a two-mode squeezing interaction through an OPA
before the photodetection. Here one evaluates the sum $D_{+}(\phi) = N_1(\phi) + N_2(\phi)$
between the detected photons in order to retrieve the information about $\phi$.
(a) Minimised sensitivity $\sens_{1\, {\rm min}}$ of the interferometer a function of $N_{\rm tot}$.
The optimal input states are two squeezed vacua with different energy (for $N_{\rm tot} \ll 1$)
or a squeezed coherent state and the vacuum (for $N_{\rm tot} \gg 1$).
The shot-noise and the Heisenberg scalings are also reported for comparisons.
(b) Minimised sensitivity $\sens_{\eta\, {\rm min}}$  as a function of $r_1$ for different
values of $N_{\rm tot}$. (c) Minimised sensitivity $\sens_{\eta\, {\rm min}}$
as a function of $r_1$ for different values of $N_{\rm tot}$ and $\eta$. We can see
that as $r_1$ increases $\sens_{\eta\, {\rm min}}$ becomes independent of
$\eta$ and depends only on the total energy $N_{\rm tot}$.}\label{f:passive:active}
\end{figure}
As we have seen, in the presence of losses ({\it i.e.} $\eta < 1$) the Heisenberg scaling is no longer
achievable by the passive-passive configuration. Nevertheless, if we substitute to the second
beam splitter an OPA (characterised by the squeezing parameter $r_1$),
thus we are using an active detection stage, we can face the losses. Moreover, upon a
suitable optimisation with respect to the squeezing parameter $r_1$
we can still obtain the Heisenberg scaling. This scenario is shown in fig.~\ref{f:passive:active}.
We note that the use of an active detection stage based on an OPA requires
to measure the sum
\begin{eqnletter}
D_{+}(\phi) &\equiv& \hbox{Tr}[\varrho_{{\rm in}} (N_c + N_d)]\,, \\
&=& N_1(\phi) + N_{2}(\phi)\,,
\end{eqnletter}
since the difference is now independent of $\phi$, as the reader can easily verify.
Therefore, the sensitivity is defined as:
\begin{equation}
\sens_{1}(\phi) = \frac{\sqrt{\hbox{var}[D_{+}(\phi)]}}{|\partial_\phi D_{+}(\phi)|}\,.
\end{equation}
\par
Assuming $\eta = 1$, the optimisation shows that we identify two regimes:
for low energy ($N_{\rm tot} \ll 1$) the optimal input
state is a couple of squeezed states with different energy, but here we can reach only the shot-noise scaling;
for $N_{\rm tot} \gg 1$, however, the use of a squeezed coherent state together with the vacuum
allows to reach the Heisenberg scaling, namely (details can be found in ref.~\cite{spara:PRA}):
\begin{equation}
\sens_{1\, {\rm min}} \approx \frac{1+\sqrt{2}}{\sqrt{2} N_{\rm tot}}\,.
\end{equation}
The result is shown in fig.~\ref{f:passive:active}~(a).
In fig.~\ref{f:passive:active}~(b) we plot $\sens_{1\, {\rm min}}$ as a function of $r_1$. As we have
mentioned above, the use of an active detection can fight the detrimental effects of losses
({\it i.e.} non-unit quantum efficiency) on
the sensitivity. This can be seen analysing fig.~\ref{f:passive:active}~(c): here we plot $\sens_{\eta\, {\rm min}}$
as a function of $r_1$ for different values of the quantum efficiency $\eta$ and $N_{\rm tot}$.
It is clear that, after the optimisation, as $r_1$ increases we have
\begin{equation}
\sens_{\eta\, {\rm min}} \to \sens_{1\, {\rm min}}\,,
\end{equation}
that is the Heisenberg scaling is restored.
\par
\begin{figure}[h!]
\begin{center}
\includegraphics[width=0.95\textwidth]{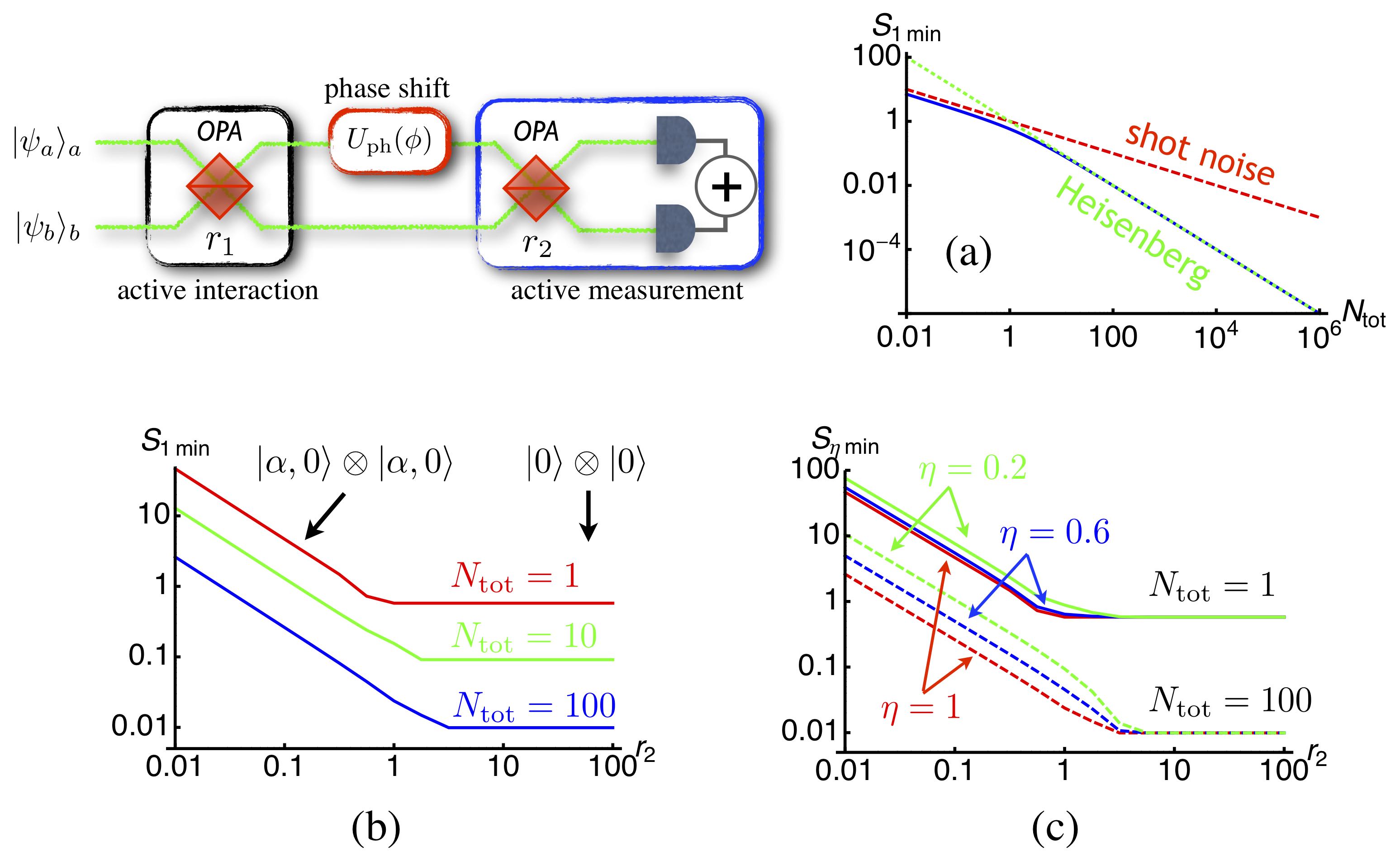}
\end{center}
\vspace{-0.5cm}
\caption{Interferometer with active interaction and detection stage:
the two input states interact at a first OPA before the phase
shift and, then, they undergo another two-mode squeezing interaction through a
second OPA before the photodetection. The sum $D_{+}(\phi) = N_1(\phi) + N_2(\phi)$
between the detected photons is then evaluated in order to retrieve the information about $\phi$.
(a) Minimised sensitivity $\sens_{1\, {\rm min}}$ of the interferometer a function of $N_{\rm tot}$.
The shot-noise and the Heisenberg scalings are also reported for comparisons.
(b) Minimised sensitivity $\sens_{\eta\, {\rm, min}}$  as a function of $r_2$ for different
values of $N_{\rm tot}$ (we recall that we have performed the optimisation over all the
other involved parameters \cite{spara:PRA}). The optimal input states, are two coherent states (for $r_{2} \ll 1$) or the vacuum (for $r_{2} \gg 1$). (c) Minimised sensitivity $\sens_{\eta\, {\rm min}}$
as a function of $r_2$ for different values of $N_{\rm tot}$ and $\eta$. We can see
that as $r_2$ increases $\sens_{\eta\, {\rm min}}$ becomes independent of
$\eta$ and depends only on the total energy $N_{\rm tot}$.}\label{f:active:active}
\end{figure}
We now turn our attention to active interferometers, in which the first beam splitter of the
passive ones is substituted with an OPA with squeezing parameter $r_1$.
The study of this kind of interferometers shows that the use of a passive
detection stage does not allow to go beyond the shot-noise limit, also in the
lossless case \cite{spara:PRA}. Therefore, here we consider only the configuration
with an active detection stage characterised by an OPA with squeezing parameter $r_2$
(see the scheme in fig.~\ref{f:active:active}). Moreover, the numerical analysis proves that the sensitivity
is minimised for $r_2 \gg 1$ \cite{spara:PRA}, thus we consider this regime. The result for
$\eta = 1$ is reported in fig.~\ref{f:active:active}~(a), where we report $\sens_{1\, {\rm min}}$
as a function of $N_{\rm tot}$: also in this case the Heisenberg scaling is achieved as
the energy increases. Figure~\ref{f:active:active}~(b) shows $\sens_{1\, {\rm min}}$ as a
function of $r_2$: it is clear that the minimum of the sensitivity is obtained when
the OPA employed at the detection introduce a large number of photons ($r_2 \gg 1$),
as mentioned above. It is interesting to note that the optimal input states
depend on the energy added by the active detection stage: if $r_2 \ll 1$ then we should
use two coherent states with the same energy, whereas for $r_2 \gg 1$ the
best choice is the vacuum state.
\par
Finally, as in the passive/active interferometer, in fig.~\ref{f:active:active}~(c)
we plot $\sens_{\eta\, {\rm min}}$ as a function of $r_2$ and different values of $N_{\rm tot}$
and $\eta$: the use of an active detection stage allows to reach the Heisenberg scaling
of the lossless case also in the presence of non-unit quantum efficiency, namely
(see ref.~\cite{spara:PRA}):
\begin{equation}
\sens_{1\, {\rm min}} \approx \frac{1}{\sqrt{N_{\rm tot}(N_{\rm tot}+2)}}\,,
\end{equation}
for the large energy regime $N_{\rm tot} \gg 1$.


\section{Practical quantum illumination}\label{s:IMAG:qill}

In sections~\ref{s:QINT} and \ref{s:bounds} we have seen that the quantum properties of light, such as the
``squeezing'', can be a useful resource to enhance the sensitivity of interferometers
going beyond the quantum shot-noise limit and finding applications in gravitational
wave antennas \cite{grav:11,DDR:15}.
\par
Here we focus on the use of quantum correlation and, in particular,  the entanglement
existing between two beams of light
generated by an OPA, the so-called twin beams (see appendix~\ref{app:TWB}),
in order to detect the presence of an opaque object
in a very noisy background.
\par
We will describe the object as a beam splitter embedded in a background of ``thermal'' radiation:
the problem is then
to distinguish the photons scattered by the object from those belonging to the background,
when the latter is predominant.
This scheme is known as \emph{quantum illumination} \cite{lloyd:08,tan:08}.
From the theoretical point of view, one can find an optimal strategy
based on twin beams which outperforms any classical one \cite{tan:08,sha:NJP:09}. However,
this kind of receivers turns out to be very challenging from the experimental point
of view \cite{sha:NJP:09,sha:PRA:09}.
\par
In the following we show the main results
one can obtain using a quantum illumination protocol based on current technology
\cite{ivano:rev,INRIM:10} which performs astonishingly better than a classical one based on
classically correlated light \cite{INRIM:13}. As a matter of fact, this protocol cannot aim to achieve
the optimal target-detection bounds of ref.~\cite{tan:08}, but it exhibits
very large quantum enhancement and robustness against noise as in the case
of the the original idea. 
\par
\begin{figure}[h!]
\begin{center}
\includegraphics[width=0.98\textwidth]{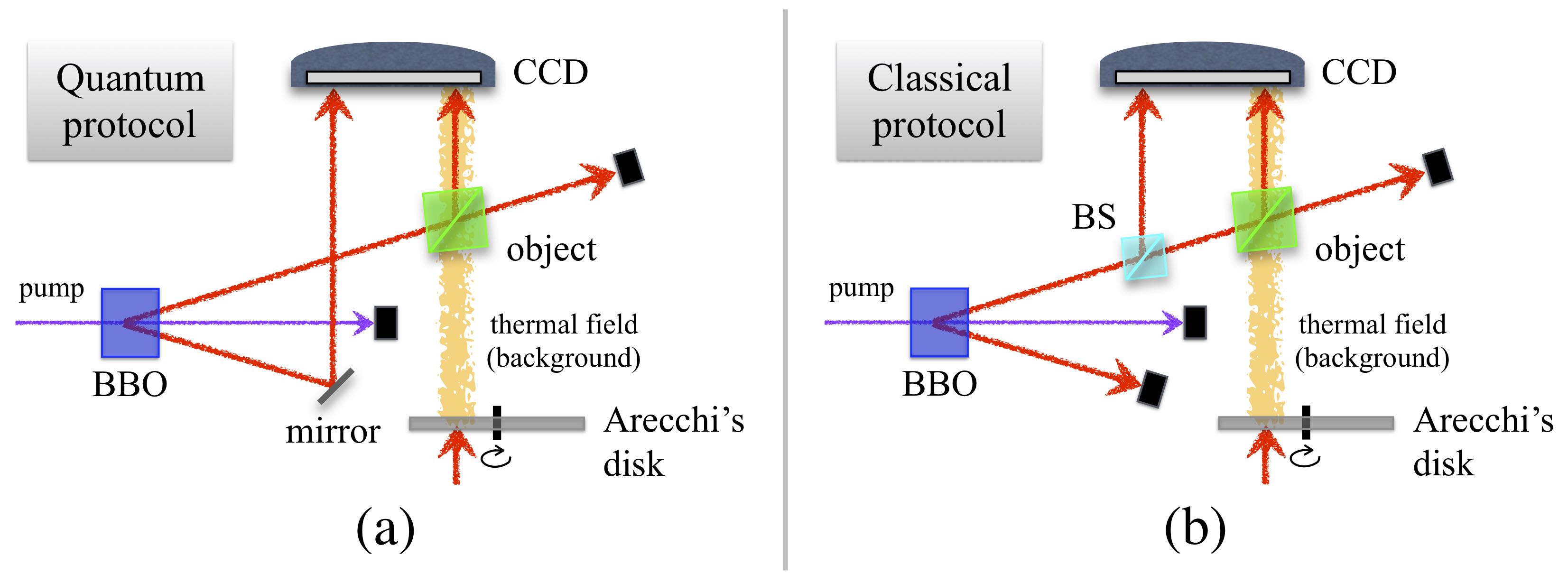}
\end{center}
\vspace{-0.5cm}
\caption{Schematic view of the quantum illumination protocol.
(a)~Quantum illumination protocol: one of the two beams of a twin-beam state
generated by a BBO is reflected
toward the detection system. If the beam splitter (the object)
is present, the correlated beam is partially detected by means of
a CCD camera, together with
the thermal field from the Arecchi's disk; otherwise it
is lost. (b)~Classical illumination protocol: now one beam 
of the twin beams is stopped and the other one is split at a beam splitter (BS)
for generating correlated multithermal beams. The pump power is adjusted in order to
obtain the same energy as in the case (a). Further details about the experimental
setup can be found in ref.~\cite{INRIM:13}.}\label{f:QI:scheme}
\end{figure}
In fig.~\ref{f:QI:scheme} we report a pictorial view of the experimental schemes
used to implement the quantum illumination protocol and the classical counterpart.
The twin beams are generated by a beta-barium borate crystal (BBO)
with an average number of photons per spatio-temporal mode $\mu = 0.075$.
They are then addressed to a high quantum efficiency CCD camera, able to measure the correlations
between the two beams. In the quantum illumination protocol, see fig.~\ref{f:QI:scheme}~(a),
an object, a 50:50 beam splitter, intercepts one of the two correlated beams
and superimposes to it a thermal background produced by scattering a laser
beam on an Arecchi's rotating ground glass. We assume that the background is
caracterised by an average number of photons $N_b$ distributed among $M_b$
modes. The other beam is instead directly detected.
\par
When the beam splitter is removed, only the background reaches the detector.
In the classical illumination protocol, reported in fig.~\ref{f:QI:scheme}~(b),
we should substitute to the twin beams two classical correlated beams.
These latter are obtained by splitting a single beam from the BBO but
adjusting the pump power in order to have the same intensity, time
and spatial coherence properties as in the quantum case \cite{INRIM:13}.
As shown in appendix~\ref{app:TWB}, the single beam of a twin beam
corresponds to a thermal state: when the thermal state is split at a beam splitter
the two emerging beams are classically correlated.
\par
Whereas the reader can find a thorough and detailed analysis of the source of the
quantum enhancement in refs.~\cite{INRIM:13}, \cite{INRIM:14} and \cite{ragy:14},
here we only report the results about the error probability in the discrimination
of the presence from the absence of the object.
The figure of merit we used in our experiment is the correlation in the
photon numbers $N_1$ and $N_2$ detected by pairs of pixels of the CCD
intercepting correlated modes of beams $a_1$and $a_2$, respectively.
In particular we focus on the covariance:
\begin{equation}
\Delta_{1,2} = \langle N_1 N_2 \rangle - \langle N_1 \rangle \langle N_2 \rangle\,,
\end{equation}
where $\langle \cdots \rangle$ is the average over the set of
a given number of realisations (since the experiment uses a CCD camera,
the number of realisations is the number of correlated pixels pairs) \cite{INRIM:13}
and $N_k = a_k^{\dag}a_k$, $k=1,2$.
\par
\begin{figure}[h!]
\begin{center}
\includegraphics[width=0.95\textwidth]{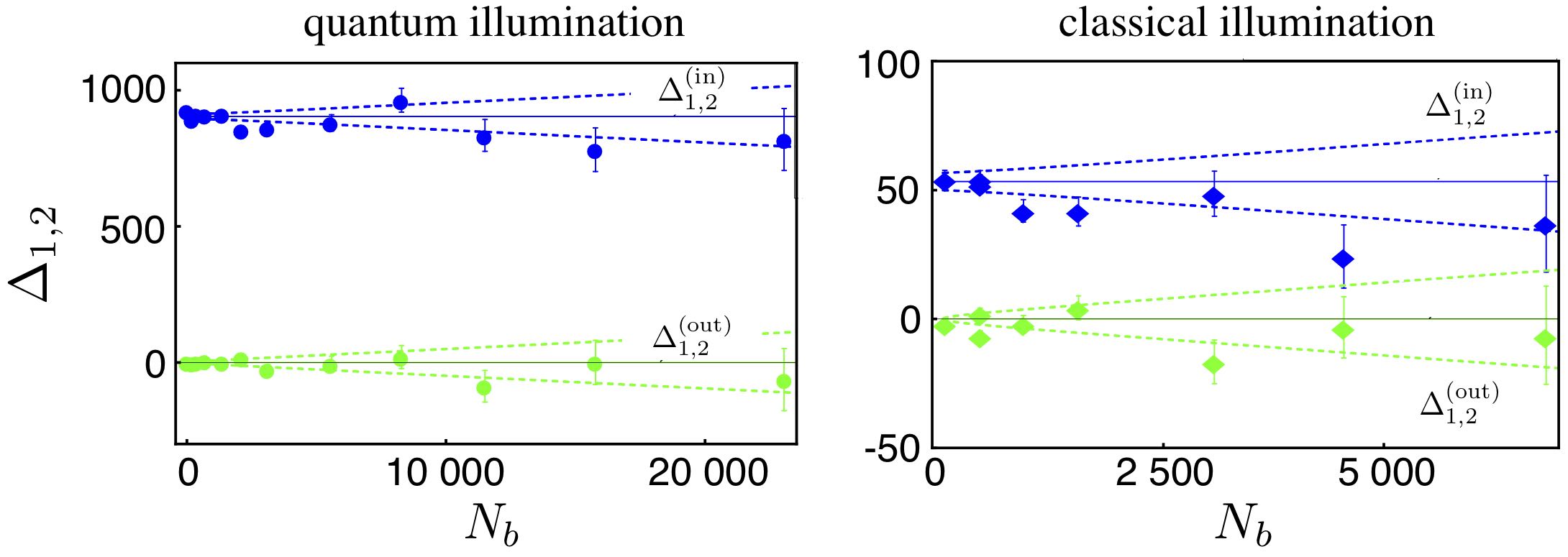}
\end{center}
\vspace{-0.5cm}
\caption{Typical plot of the covariance in the presence, $\Delta_{1,2}^{\rm (in)}$
(blue), and in the absence, $\Delta_{1,2}^{\rm (out)}$ (green), of the object
as a function of the average number of background photons $N_b$
distributed among $M_b = 1300$ modes. The plot on the left
refers to the quantum illumination protocol, whereas the one on the
right to the classical counterpart as described in the text. The bars
are the effect of the background noise on the covariance estimation.
We also report the expectations (horizontal lines) and the
corresponding uncertainty interval as calculated theoretically.
It is clear that the presence or the absence of the object can be
inferred by addressing the value of the covariance. Notice
the enhancement due to the quantum protocol. Data comes from
ref.~\cite{INRIM:13}.}\label{f:cov:QI:CI}
\end{figure}
The problem of discriminating the presence from the absence of the object
is then equivalent to that of distinguishing between two
corresponding values of $\Delta_{1,2}$.
More precisely, we can choose a threshold value of
$\Delta_{1,2}$, above which we can infer that the object is present
otherwise it is not. Typical results are shown in
fig.~\ref{f:cov:QI:CI}, where the measured values of the covariance
and their error bars are
plotted as a function of the background photons $N_b$ for the
quantum illumination (left panel) and classical illumination (right panel):
note how the quantum protocol outperforms the classical one.
\par
\begin{figure}[h!]
\begin{center}
\includegraphics[width=0.55\textwidth]{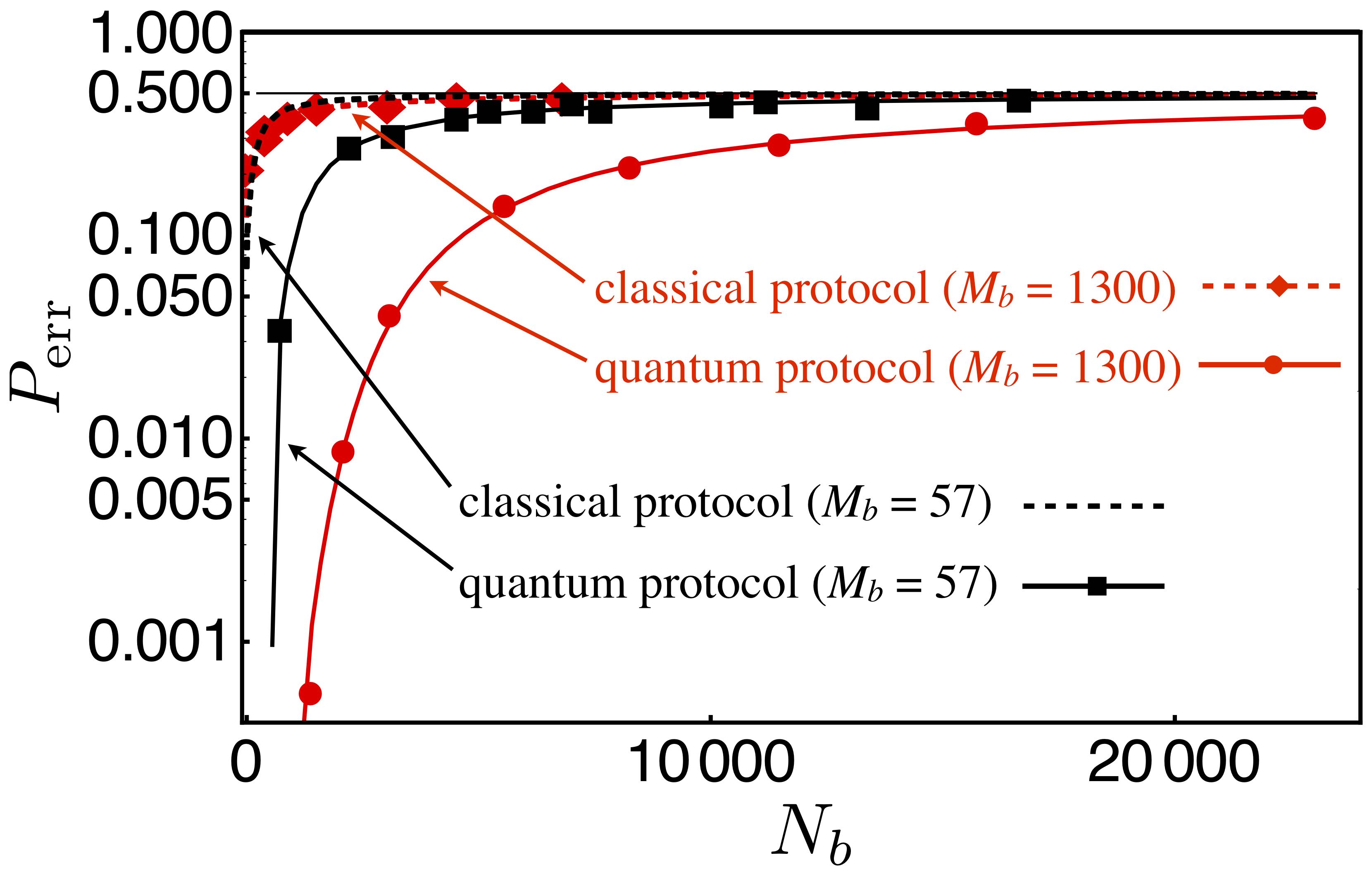}
\end{center}
\vspace{-0.3cm}
\caption{Plot of the target detection error probability as a function of
the background photons $N_b$ and for two values of the number of
modes, $M_b = 57$ (black) and $M_b = 1300$ (red).
The symbols refer to the experimental data whereas the lines are
the theoretical previsions. The points corresponding to $M_b = 1300$
have been obtained by processing the data in fig.~\ref{f:cov:QI:CI} as described
in the text.}\label{f:QI:Perr}
\end{figure}

Since the experimental values unavoidably fluctuates,
we have to evaluate the error probability $P_{\rm err}$
of the discrimination, that is the probability to infer the wrong answer given
a value of the covariance. If we know only the mean value and the variance
of a certain quantity,  we should assign a Gaussian probability distribution to
that quantity according to the maximum entropy principle (see, for instance,
ref.~\cite{oli:metro}). Therefore it is possible to associate a probability
distribution with each point of fig.~\ref{f:cov:QI:CI} and we can evaluate the
error probability by fixing a threshold value of the covariance
chosen in order to minimise $P_{\rm err}$.
In fig.~\ref{f:QI:Perr} we show $P_{\rm err}$ as a function of $N_b$ for two values of the
number of background modes $M_b$, in the case of both the quantum illumination
protocol and the classical one. We can also see that the quantum improvement becomes
larger as $M_b$ increases. In particular, the points corresponding to $M_b = 1300$
have been obtained by processing the data in fig.~\ref{f:cov:QI:CI} as described above.
\par
Entanglement is usually a fragile resource which should be
carefully protected by the environment and, thus,
the advantages of entangled and quantum states
can be exploited only in advanced quantum laboratories
or for academic discussions. Nevertheless, the results we presented in this section
show that it is possible to obtain orders of magnitude improvements
compared to the classical protocol, independent of the amount of noise
and losses, by using devices available nowadays. Therefore,
this kind of ``practical'' quantum illumination protocol based on
photon counting can have interesting potentialities
to promote the usage of quantum correlated light in real and more complex
scenarios \cite{microw:QI,ray:QI}.


\section{Beyond single-interferometer setups}\label{s:coupled:qgrav}

In section~\ref{s:bounds} we have seen how quantum light can be exploited in
interferometers to improve the estimation of an unknown phase shift
also beating the shot-noise limit and, thus, opening the way to a new
generation of quantum-enhanced interfereometers \cite{grav:11}.
However, as illustrated in section~\ref{s:IMAG:qill}, quantum correlations
between light beams can be used to successfully outperform
the performance of standard imaging protocols. The question that
now arises is whether it is possible to exploit quantum correlations
to further improve interferometers. In fact,  recently quantum correlations
have received a lot of attention as a key ingredient in advanced
quantum metrology protocols \cite{adv:QM,microscopy}.
\par
In the following section we will review the main theoretical
results we obtained coupling two interferometers via quantum-correlated
beams. In particular, we show how protocols based on quantum-correlated
interferometers lead to substantial advantages with respect to the use of
classical light, up to a noise-free scenario for the ideal lossless case
\cite{coupled:13,coupled:15}
also in the presence of some additional noise \cite{benatti:17}.


\section{Probing the noncommutativity of position and momentum}\label{s:probing}

One of the most interesting applications of interference devices concerns
the quantum gravity tests. The non-commutativity at the Planck scale
($l_p = 1.616 \times 10^{-35}~m$) of position variables in different directions
gives rise to quantum fluctuations of the space geometry
\cite{aschieri:09,aschieri:10} which, under particular
conditions, could lead to detectable effects in cavities with microresonators
\cite{aspel:12} or in two coupled interferometers, the so-called holometer
\cite{hogan:12,holo:16}. In the latter case, the predicted non commutativity
leads to an additional phase noise, the ``holographic noise''.
\par
\begin{figure}[h!]
\begin{center}
\includegraphics[width=0.3\textwidth]{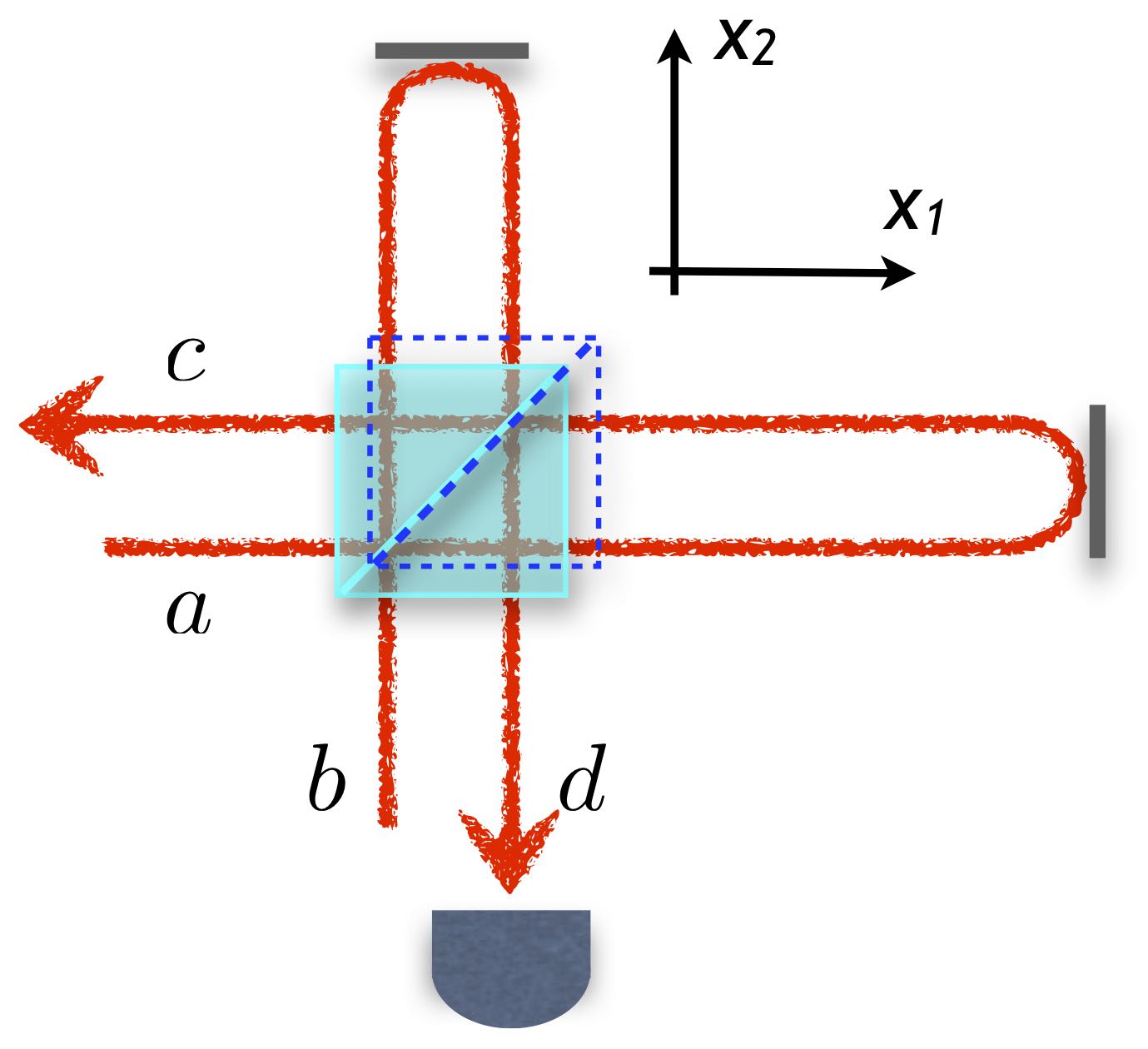}
\end{center}
\vspace{-0.3cm}
\caption{The jitter in the beam splitter position $(x_1,x_2)$ leads to fluctuations in the
measured phase. See the text for details.}\label{f:michelson:jitter}
\end{figure}
If we consider a single interferometer based on a single beam splitter,
such as the Michelson interferometer sketched in fig.~\ref{f:michelson:jitter},
in the presence of the holographic noise
the jitter in the beam splitter position leads to fluctuations of the measured phase.
Since, the phase shift can be seen as a measurement of the beam splitter
position, the phase fluctuations are directly related to the fluctuations $\Delta x_k^2 = l_p \, L$
of the coordinate $x_k$, $k=1,2,3$, $L$ being the length of the interferometer's arm
(for the sake of clarity we recall that $L = 600$~m for GEO600 and
$L = 4$~km for LIGO) \cite{hogan:12}.
 Therefore, the holographic noise accumulates as a random walk and becomes detectable.
\par
Unfortunately, the most precise interferometers able to detect gravitational waves have
a resolution at low frequencies $f \ll c/(4\pi L)$ not enough to detect the holographic noise
(the interested reader can find plenty
of details about the holographic noise and its origin in ref.~\cite{hogan:12}).
Nevertheless, it could be possible to identify this noise by evaluating the cross-correlation 
between the two equal interferometers of the holometer when placed in the same
space-time volume: while the shot noises of the interferometers are uncorrelated
and do vanish over a long integration time, the holographic noises are not. Moreover,
one can also ``turning off'' the holographic fluctuations by separating the space-time
volumes, thus obtaining a background estimation. As a matter of fact, the
ultimate limit for the holometer sensibility is related to the shot noise.
In the following we will discuss how it could be possible to go beyond this limit
by using the quantum optical states we introduced in section~\ref{s:bounds}
in the case of a single interferometer.
\par
\begin{figure}[h!]
\begin{center}
\includegraphics[width=0.45\textwidth]{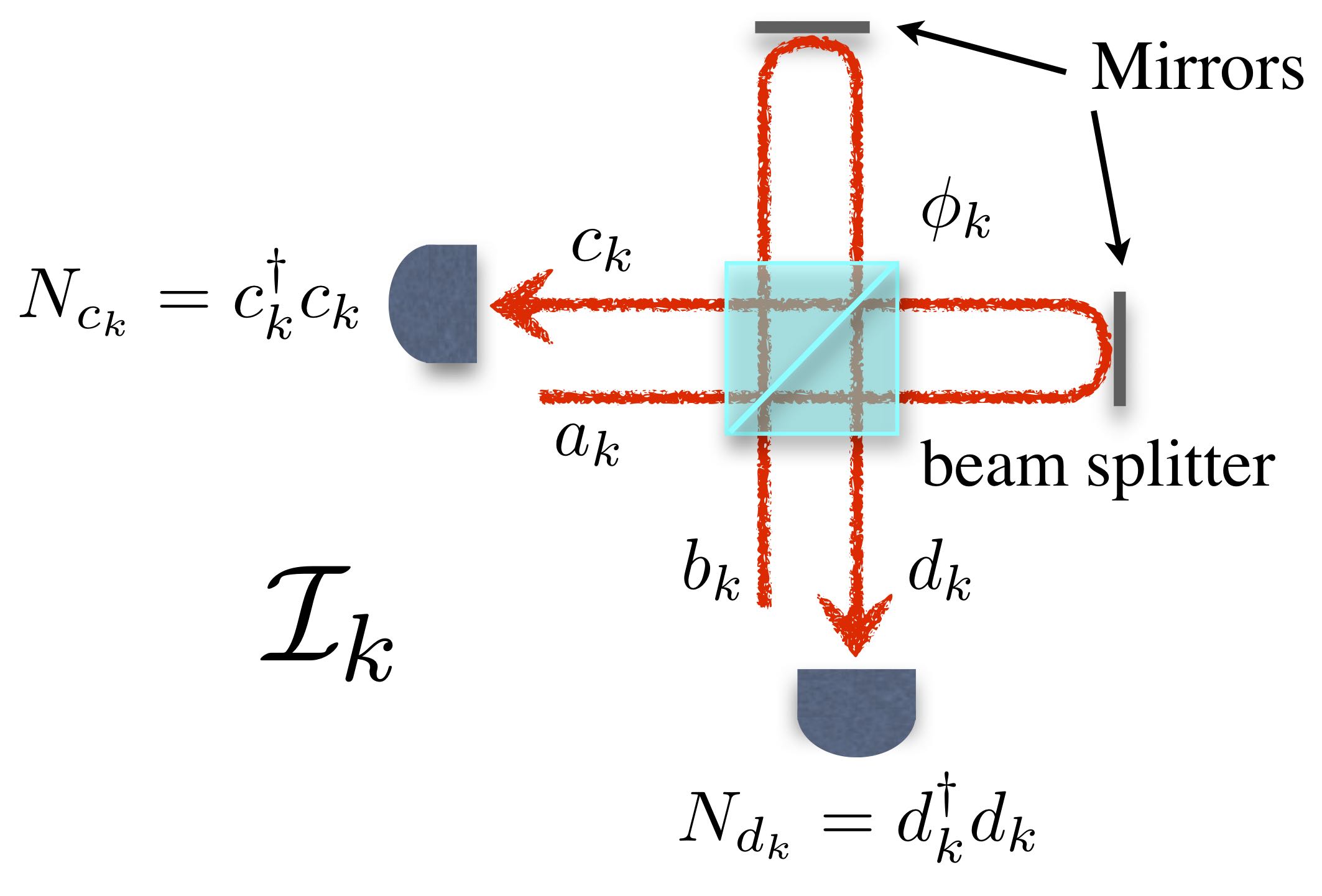}
\end{center}
\vspace{-0.3cm}
\caption{Scheme of the Michelson interferometer ${\cal I}_k$, $k=1,2$.
We reported the involved modes and the measured quantities.}\label{f:michelson}
\end{figure}
We consider two Michelson interferometers ${\cal I}_k$, $k=1,2$, as shown in fig.~\ref{f:michelson},
where we reported the involved modes and the measured number operators at
the outputs.
The $k$-th interferometer detects the phase shift $\phi_k$.
We are interested in a joint measurement, thus we should consider a suitable
operator $C(\phi_1,\phi_2)$ with expectation
\begin{equation}
\langle C(\phi_1,\phi_2) \rangle =
\hbox{Tr}[\varrho_{12}\,C(\phi_1,\phi_2)]\,,
\end{equation}
where $\varrho_{12}$ is the
overall density matrix associated with the four-mode state of the light beams injected
into the two interferometers ${\cal I}_1$ and ${\cal I}_2$.
\par
\begin{figure}[h!]
\begin{center}
\includegraphics[width=0.75\textwidth]{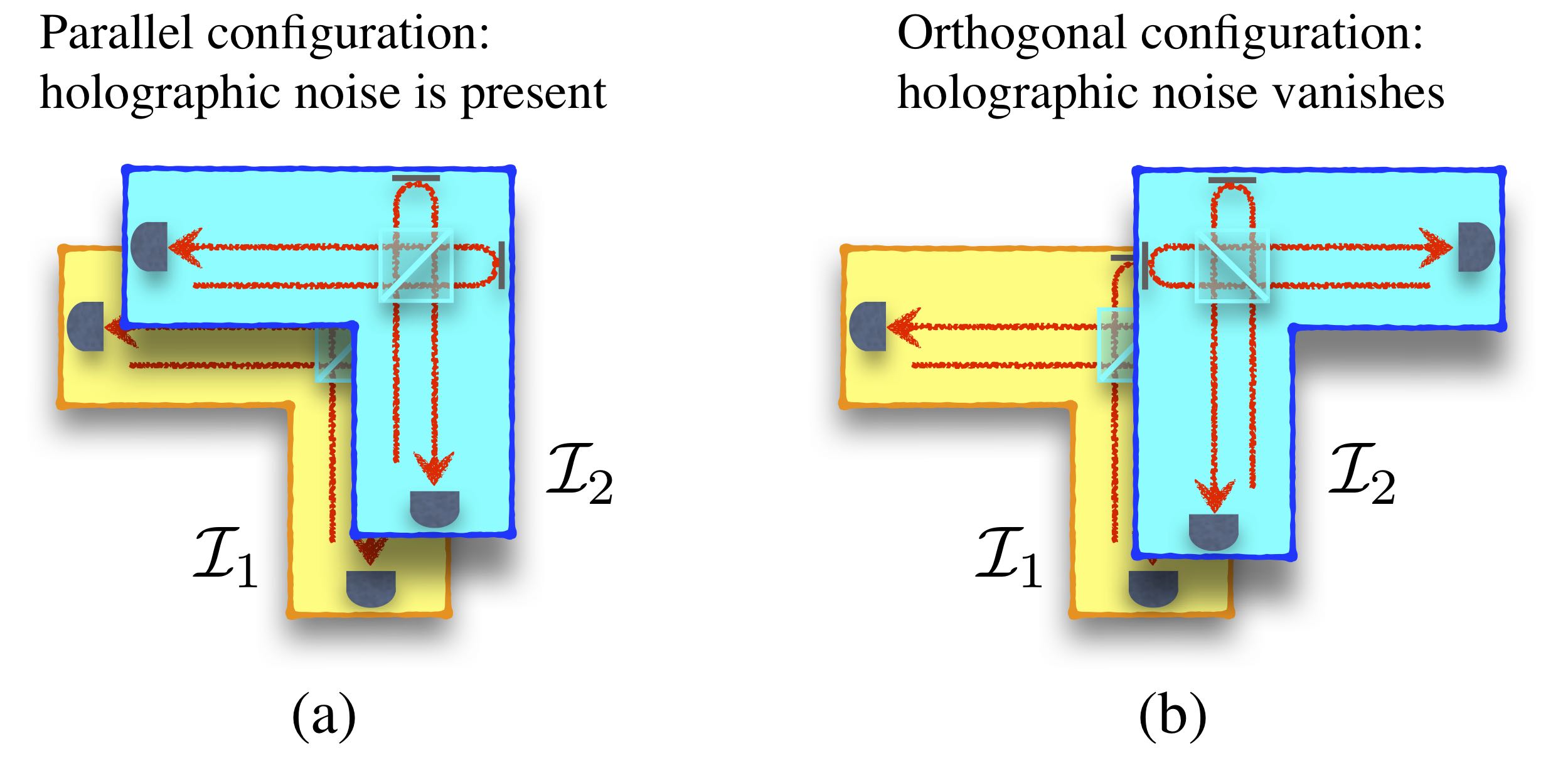}
\end{center}
\vspace{-0.5cm}
\caption{Two possible configurations of the two Michelson interferometers.
(a) In the parallel configuration the interferometers are in overlapping
space-time volumes: the holographic noise correlates the outputs.
(b) In the orthogonal configuration the space-time volumes of the
interferometers are no longer overlapping and the correlation induced
by the holographic noise vanishes.}\label{f:par:orth}
\end{figure}

The two configurations used to observe the holographic noise are depicted in
fig.~\ref{f:par:orth}. Figure~\ref{f:par:orth}~(a) shows the parallel configuration, in which
the two interferometers occupy the same space-time volume: in this case the
holographic noise induces the same fluctuation on the phase shift, thus leading
to a correlation of the intensity fringes. When the interferometers are in the
orthogonal configuration, fig.~\ref{f:par:orth}~(b), their space-time
volumes are no longer overlapping and the correlation induced by the
holographic noise vanishes: this can be used as a reference, that is
a measurement of the ``background''.
\par
Following ref.~\cite{coupled:13}, we can describe the statistical
properties of the phase shift fluctuations induced by the holographic
noise by means of a suitable probability density function
$f_{x}(\phi_1,\phi_2)$, $x=\parallel,\perp$.
Of course, since if we address the single interferometer we cannot distinguish
between the two configurations, given the marginals:
\begin{equation}
{\cal F}^{(k)}_x(\phi_k) = \int f_x(\phi_1,\phi_2)\, d\phi_h\,,
\end{equation}
with $h,k=1,2$ and $h \ne k$, we should have:
\begin{equation}
{\cal F}^{(k)}_{\parallel}(\phi_k) = {\cal F}^{(k)}_{\perp}(\phi_k)\,;
\end{equation}
on the other hand, since in the orthogonal configuration there are not correlations between the two
interferometers, we should also require:
\begin{equation}
f_{\perp}(\phi_1,\phi_2) = {\cal F}^{(1)}_{\parallel}(\phi_1)\,{\cal F}^{(2)}_{\perp}(\phi_2)\,.
\end{equation}
Therefore, the actual expectations of an operator ${\cal O}(\phi_1,\phi_2)$,
which depends on the phase shifts, should be averaged over $f_x(\phi_1,\phi_2)$, namely:
\begin{equation}
\langle {\cal O}(\phi_1,\phi_2) \rangle \to
{\cal E}_x [{\cal O}(\phi_1,\phi_2)] = \int f_x(\phi_1,\phi_2)\, \langle {\cal O}(\phi_1,\phi_2) \rangle
\, d\phi_1 \, d\phi_2\,.
\end{equation}
\par
The information about the correlation between the two phases, can be obtained
by estimating the covariance in the parallel configuration, {\it i.e.}
${\cal E}_{\parallel}[\delta\phi_1,\delta \phi_2]$, where
\begin{equation}
\delta\phi_k = \phi_k - \phi_{k,0}\,,
\end{equation}
and we introduced the central phase $\phi_{k,0}$ measured by the $k$-th interferometer.
In the limit $\delta\phi_1, \delta\phi_2 \ll 1$ we can write \cite{coupled:13}:
\begin{equation}
{\cal E}_{\parallel}[\delta\phi_1,\delta \phi_2] \approx
\frac{{\cal E}_{\parallel}[C(\phi_1,\phi_2)] - {\cal E}_{\perp}[C(\phi_1,\phi_2)]}
{\langle \partial_{\phi_1,\phi_2}^2  C(\phi_{1,0},\phi_{2,0})\rangle}\,,
\end{equation}
and we clearly see that the covariance can be estimated by measuring the
difference between the expectation values of the operator $C(\phi_1,\phi_2)$
in the two configurations.
\par
In order to observe the holographic noise, we should minimize the uncertainty
associated with the measurement of the covariance, which reads (still
in the limit $\delta\phi_1, \delta\phi_2 \ll 1$):
\begin{equation}
{\cal U}(\delta\phi_1,\delta \phi_2) \approx
\frac{\sqrt{\hbox{var}_{\parallel}[C(\phi_1,\phi_2)] + \hbox{var}_{\perp}[C(\phi_1,\phi_2)]}}
{| \langle \partial_{\phi_1,\phi_2}^2  C(\phi_{1,0},\phi_{2,0})\rangle |}\,,
\end{equation}
with $\hbox{var}_{x}[C(\phi_1,\phi_2)] = {\cal E}_{x}[C^2(\phi_1,\phi_2)] - {\cal E}_{x}[C(\phi_1,\phi_2)]^2$.
Since
\begin{equation}
\hbox{var}_{\parallel}[C(\phi_1,\phi_2)] = \hbox{var}_{\perp}[C(\phi_1,\phi_2)] = 
\hbox{var}[C(\phi_1,\phi_2)] + O(\delta\phi^2)\,,
\end{equation}
the zero-order contribution to the uncertainty reduces to:
\begin{equation}
\label{zero:unc}
{\cal U}^{\rm (0)} \approx
\frac{\sqrt{2 \, \hbox{var}_{\parallel}[C(\phi_{1,0},\phi_{2,0})]}}
{| \langle \partial_{\phi_1,\phi_2}^2  C(\phi_{1,0},\phi_{2,0})\rangle |}\,.
\end{equation}
It is worth noting that ${\cal U}^{\rm (0)}$, that is the main contribution to the
uncertainty, does not depend on the fluctuations induced by the holographic noise,
but only on the \emph{intrinsic} fluctuations due to the chosen measurement and to the
state sent through the interferometers. As we have also seen in sections~\ref{s:QINT} and \ref{s:bounds},
the choice of the measurement is strictly related to the input state and, in the following, we will
show that squeezing and entanglement can provide huge advantages in terms of
the achieved accuracy with respect to classical light \cite{coupled:13}.
\par
We start considering that the two-mode input state of the $k$-th interferometer
is exited in a coherent and a squeezed vacuum state with mean number of photons
$\mu_k$ and $\lambda_k$, respectively (see appendices~\ref{app:CS} and \ref{app:SQ}).
As discussed in section~\ref{s:bounds}, in this
case the value of $\phi_k$ can be retrieved assessing the difference $D_{k,-}(\phi_k)$
of the number of photons in the two output ports of the interferometer ${\cal I}_k$.
Therefore, we can now define:
\begin{equation}
\Delta C(\phi_1,\phi_2) \equiv \Delta D_{1,-}(\phi_1)\, \Delta D_{2,-}(\phi_2)\,,
\end{equation}
where
\begin{equation}
\Delta D_{k,-}(\phi_k) = D_{k,-}(\phi_k) - {\cal E}[D_{k,-}(\phi_k)]\,.
\end{equation}
If we assume
$\mu_1 = \mu_2 = \mu$ and $\lambda_1 = \lambda_2 = \lambda$, and consider
the optimal working regime $\phi_{1,0} = \phi_{2,0} = \pi/2$, we obtain \cite{coupled:13}:
\begin{equation}
{\cal U}^{\rm (0)}_{\rm SQ} \approx \sqrt{2}\,
\frac{\lambda + \mu (1+2\lambda-2\sqrt{\lambda+\lambda^2})}{(\lambda-\mu)^2}
\end{equation}
which, in the limit $\mu \gg \lambda \gg 1$, becomes:
\begin{equation}
{\cal U}^{\rm (0)}_{\rm SQ} \approx
\frac{1}{2\sqrt{2}\,\lambda\,\mu}\,.
\end{equation}
It is thus clear the advantage with respect to the classical case
${\cal U}^{\rm (0)}_{\rm CL} \approx \sqrt{2}/\mu$ in which only
coherent states are used. Nevertheless, in this latter case
it is worth noting that the
measurement on the covariance, involving \emph{second-order momenta},
leads to a scaling $\propto \mu^{-1}$ instead of $\propto \mu^{-1/2}$
as in the case of the single interferometer.
\par
Though the use of squeezing leads to an advantage with respect to
the classical scenario, the improvement is due to the \emph{independent}
improvement of the two single interferometers. Here, however,
we are interested in minimising the noise of the correlation between
the interferometers. It is thus natural to investigate whether the use
of quantum correlated states coupling the two interferometers
can produce further improvement.
\par
Let's now suppose to couple ${\cal I}_1$ and ${\cal I}_2$ by exciting their
input modes $a_1$ and $a_2$ (see fig.~\ref{f:michelson}) in the
twin-beam state introduced in section~\ref{s:bounds} (see also appendix~\ref{app:TWB}) which
we rewrite as::
\begin{equation}
| \hbox{TWB} \rangle\rangle_{a_1,a_2} = 
\sum_n c_n(\lambda) \, | n \rangle_{a_1} | n \rangle_{a_2}\,.
\end{equation}
where $\lambda$ is still the average number of photons of each mode
and
\begin{equation}
c_n(\lambda) = 
\sqrt{\frac{\lambda^n}{(1+\lambda)^{n+1}}}\,,
\end{equation}
Due the symmetry between the two modes we have
the peculiar property:
\begin{equation}
{}_{a_1,a_2}\langle \langle \hbox{TWB} | (a_1^\dag a_1 - a_2^\dag a_2)^M | \hbox{TWB} \rangle\rangle_{a_1,a_2} = 0
,\quad \forall M > 0.
\end{equation}
As in the previous case, we assume to send
in the other input ports of the interferometers two equal coherent states
$| \sqrt{\mu} \rangle_{b_1}| \sqrt{\mu} \rangle_{b_1}$ with average number
of photons $\mu$. Overall, we have the four-mode input state:
\begin{equation}
|\Psi \rangle = | \hbox{TWB} \rangle\rangle_{a_1,a_2}| \sqrt{\mu} \rangle_{b_1}| \sqrt{\mu} \rangle_{b_1}\,.
\end{equation}
If we choose the working regime $\phi_{1,0} = \phi_{2,0} = 0$ and,
of course, in the absence of the holographic noise, the
two interferometers of the holometer behave like two \emph{completely transparent
media}, as one can easily check from the input-output relations (\ref{in:out:int}):
the output modes $c_1$ and $c_2$, coming from $a_1$ and $a_2$, respectively,
exhibit \emph{perfect} correlation between the number of photons.
\par
If we now define the observable
\begin{equation}
C(\phi_1,\phi_2) = \Delta^2 (N_{c_1} - N_{c_2})\,,
\end{equation}
where $N_{c_k} = c_k^\dag c_k$, we have $\hbox{var}[C(\phi_1,\phi_2)] = 0$ in
eq.~(\ref{zero:unc}), whereas the denominator reads:
\begin{equation}
| \langle \Psi |\partial_{\phi_1,\phi_2}^2  C(\phi_{1,0},\phi_{2,0}) |\Psi \rangle | =
\frac12 \, \mu \sqrt{\lambda +\lambda^2}\,.
\end{equation}
Therefore, if $\mu,\lambda \ne 0$ we find the striking result:
\begin{equation}
{\cal U}^{(0)}_{\rm TWB} = 0\,.
\end{equation}
Remarkably, the perfect correlation existing between
the two beams leads to a vanishing zero-order contribution to the uncertainty
of the covariance. In this scenario any very faint perturbation which correlates the
interferometers can be detected, since it cannot be masked by a residual noise.
We note that in the presence of fluctuations due to the holographic noise,
a little portion of the coherent states is reflected to the monitored ports, thus
guaranteeing the sensitivity to the covariance of the holographic noise phase shift.
\par
\begin{figure}[h!]
\begin{center}
\includegraphics[width=0.55\textwidth]{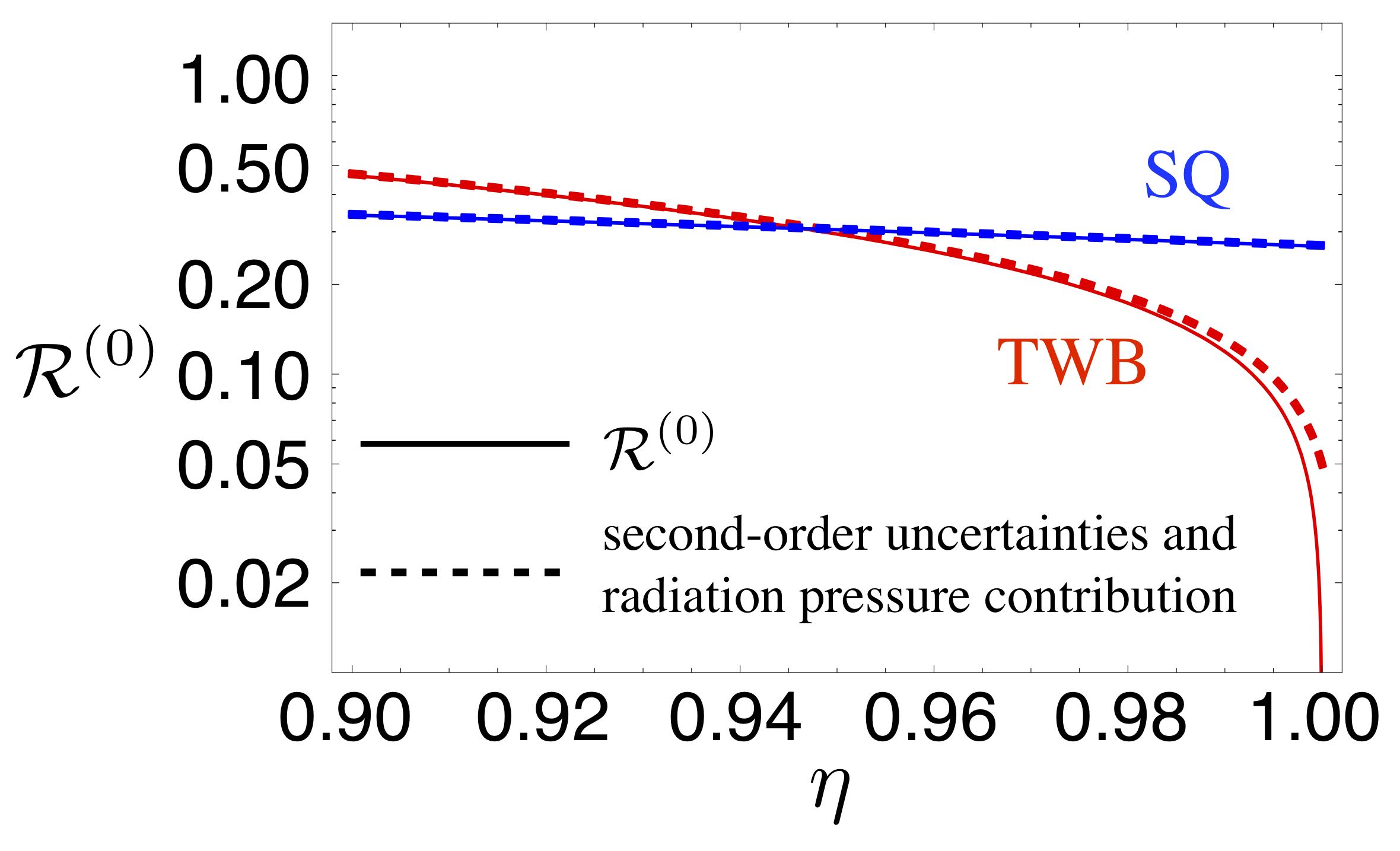}
\end{center}
\vspace{-0.5cm}
\caption{Plot of the ratio ${\cal R}^{(0)} = {\cal U}^{(0)}/{\cal U}^{(0)}_{\rm CL}$ as a function of
the overall transmission-detection efficiency $\eta$ for the squeezed vacua and
the twin beams (solid lines). We consider the realistic values $\mu = 2 \times 10^{23}$ for the
coherent state intensities and $\lambda = 0.5$ for the squeezed vacua and each of
the twin beams. The dashed lines refer to the second order uncertainties where
we also added the radiation pressure contribution. Here we considered
a mirror mass $10^2$~kg and a central light frequency
$\omega = 3.14\times 10^{15}$~Hz (corresponding to a wavelength of
600~nm) and a measurement time of $10^{-3}$~s.
Adapted from Ref.~\cite{coupled:13}.}\label{f:ratio:unc}
\end{figure}
As a matter of fact, in real experiments we should also consider other effects
which may affect the ideal results obtained above. One of the main contribution
which can increases the uncertainty of the covariance measurement is a
non-unit overall transmission-detection efficiency $\eta$ (see appendix~\ref{app:eta}).
In fig.~\ref{f:ratio:unc} shows the ratio
\begin{equation}
{\cal R}^{(0)} = \frac{{\cal U}^{(0)}}{{\cal U}^{(0)}_{\rm CL}}
\end{equation}
as a function of $\eta$
for the squeezed vacua and the twin beams (solid lines). We can see that
there exists a threshold on $\eta$ (that depends on the other involved parameters) above which the
measurement involving the twin beams outperforms both the classical strategy
and the one based on uncorrelated squeezed vacua. In this region of
high efficiency we have a significant reduction of the uncertainty. Moreover,
our analysis proves that also below that threshold nonlcassical squeezed light
beats the performance of classical coherent light. In the limit $\mu \gg 1$ and
$\lambda \ll 1$ (as considered in fig.~\ref{f:ratio:unc}), we find:
\begin{equation}
{\cal R}^{(0)}_{\rm SQ} = \frac{{\cal U}^{(0)}_{\rm SQ}}{{\cal U}^{(0)}_{\rm CL}}
\approx 1-2\eta\sqrt{\lambda}\,,\qquad (\mu \gg 1, \lambda \ll 1)
\end{equation}
whereas
\begin{equation}
{\cal R}^{(0)}_{\rm TWB} = \frac{{\cal U}^{(0)}_{\rm TWB}}{{\cal U}^{(0)}_{\rm CL}}
\approx \sqrt{\frac{2(1-\eta)}{\eta}}\,,\qquad (\mu \gg 1, \lambda \ll 1)
\end{equation}
and an improvement with respect the classical case can be obtained for
$\eta > 2/3 \approx 0.67$. On the other hand, if $\mu \gg \lambda \gg 1$ we obtain
\begin{equation}
{\cal R}^{(0)}_{\rm SQ} \approx (1-\eta)-\frac{\eta}{4\lambda}\,,\qquad
(\mu \gg \lambda \gg 1)
\end{equation}
and
\begin{equation}
{\cal R}^{(0)}_{\rm TWB} \approx 2\sqrt{5}(1-\eta)\,. \qquad
(\mu \gg \lambda \gg 1)
\end{equation}
Therefore, also in this case we the approach based on twin beams outperforms
the classical one (for $\eta > 0.776$) and the uncertainty drop to zero as
$\eta \to 1$.
\par
For the sake of completeness, in fig.~\ref{f:ratio:unc} we also plot the uncertainty
reduction normalised to ${\cal U}^{(0)}_{\rm CL}$ as a function of $\eta$
(dashed lines) considering the second-order uncertainties and the radiation
pressure contribution \cite{caves:81}.
We recall that in the case of a single interferometer fed by squeezed light,
the amplitude of the noise due to the radiation pressure decreases as the squeezing
parameter increases. In the present case we have an analogue behaviour.
If we consider reasonable values of the involved parameters, the radiation pressure noise
is completely negligible.
We can see in fig.~\ref{f:ratio:unc} that for the realistic parameters
we have chosen we have just a very small correction.
The reader can find further details about this last point in ref.~\cite{coupled:13}.
\par
Remarkably, our results not only demonstrate that the use of quantum and entangled
states of light allows reaching much higher sensibility for the realisation of experiments
to test quantum gravity, but also put forward new opportunities for the
design of innovative interferometric schemes, as we will discuss in the next section.


\section{Squeezed and entangled light in correlated interferometry}\label{s:sq:corr:int}

The experimental requirements to implement the double-interferometer setup
introduced in the previous section are extremely challenging. In particular,
one should control the working regimes of the interferometers with a very
high accuracy in order to obtain the huge advantages given by twin beams.
Moreover, the quantum efficiency $\eta$ must be greater than
$\approx 0.99$ (see fig.~\ref{f:ratio:unc}).
\par
We have seen that to exploit the twin-beam properties to reduce the
uncertainty in the covariance measurement, we should choose
a regime in which the interferometers act like transparent media.
We recall that if $\phi_k$ is the measured phase, the interferometer
${\cal I}_k$ is equivalent to a beam splitter with transmissivity
$\tau_k = \cos^2(\phi_k/2)$ (the interferometer transmission).
The coherent amplitude which usually circulates in the interferometers
is so high that also a small reflection probability $1-\tau_k$ can send
to the measured ports a really high number of photons. For instance,
if only coherent states with energy $\mu$ are considered and,
therefore, the input state is
\begin{equation}
| 0 \rangle_{a_1} | 0 \rangle_{a_2} | \sqrt{\mu} \rangle_{b_1} | \sqrt{\mu} \rangle_{b_2}\,,
\end{equation} 
we have
\begin{eqnletter}
\label{COHstat}
  \langle N_{k}\rangle^{\coh}_{\eta\tau_{k}}
  &=& \langle\delta N_k ^{2}\rangle^{\coh}_{\eta\tau_{k}}
   =  \eta \mu (1-\tau_{k}), \\[1ex]
  \langle\delta N_{1} \delta N_{2}\rangle^{\coh}_{\eta\tau_{1}\tau_{2}} &\equiv& 0,
\end{eqnletter}
 where $N_k = c_k^{\dag} c_k$, $k=1,2$, $\delta {\cal O} = {\cal O} - \langle {\cal O} \rangle$,
 $\delta {\cal O}^2 = {\cal O}^2 - \langle {\cal O} \rangle^2$
and $\eta$ is the quantum efficiency of the detectors (see appendix~\ref{app:eta}).
We assume, for the sake of simplicity, that they have the same quantum efficiency.
On the other hand, when we send only the twin beams to correlate the
 interferometer the four-mode input state is
\begin{equation}
| {\rm TWB} \rangle \rangle_{a_1,a_2}  | 0 \rangle_{b_1} | 0 \rangle_{b_2}\,,
\end{equation}
and we find
\begin{eqnletter}
\label{TWBstat}
  \langle N_k \rangle^{\TWB}_{\eta\tau_{k}} &=&  \eta \tau_{k}\lambda\,, \\[1ex]
  \langle\delta N_k ^{2}\rangle^{\TWB}_{\eta\tau_{k}} &=& 
  \eta \tau_{k} \lambda (1+\eta \tau_{k} \lambda) \,,\\[1ex]
  \langle\delta N_{1} \delta N_{2}\rangle^{\TWB}_{\eta\tau_{1}\tau_{2}} &=&
  \eta^{2}\tau_{1}\tau_{2}\lambda(1+\lambda)\,,
\end{eqnletter}
where we used the same parametrization as in section~\ref{s:coupled:qgrav}.
\par
Following Ref.~\cite{coupled:15}, we introduce the noise reduction parameter
\begin{equation}
\NRF_{\pm} = \frac{\langle\delta(N_{1}\pm N_{2})^{2}\rangle}{\langle N_{1}+N_{2}\rangle}\,,
\end{equation}
that is the ratio between the variance of the sum or difference between the
detected photon numbers and the corresponding shot-noise limit.
Since a value $\NRF_{-} < 1$ is a signature of nonclassical correlations
while $\NRF_{+} < 1$ denotes anticorrelations of the photon number beyond the
classical limits, we can use $\NRF_{\pm}$ as a figure of merit for the
correlations at the output ports.
\par
\begin{figure}[h!]
\begin{center}
\includegraphics[width=0.95\textwidth]{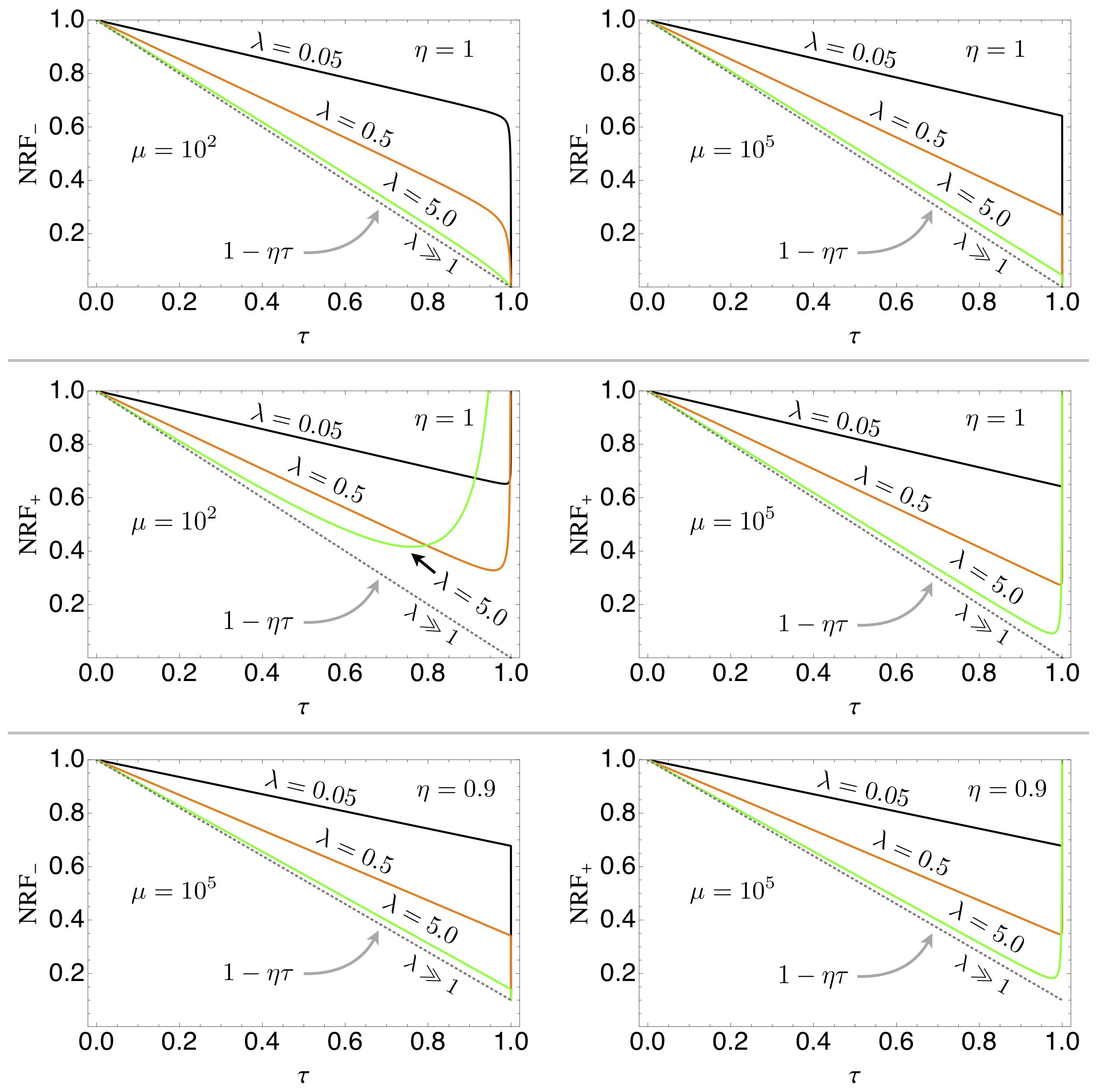}
\end{center}
\vspace{-0.5cm}
\caption{(Top) Plot of the maximised ({\it i.e.} $\psi = \pi/2$) $\NRF_{-}$
as a function of the overall interferometer
transmission $\tau$ for different values of the twin-beam energy per beam $\lambda$ and for $\mu = 10^2$
(left panel) and $\mu = 10^5$ (right panel). We set $\eta = 1$ .
(Centre) Plot of the maximised ({\it i.e.} $\psi = 0$) $\NRF_{+}$ as a function of the overall interferometer transmission $\tau$
for different values of the twin-beam energy per beam $\lambda$ and for $\mu = 10^2$
(left panel) and $\mu = 10^5$ (right panel). We set $\eta = 1$.
(Bottom) Plot of $\NRF_{\pm}$ with the same parameters as in the other plots but with $\eta = 0.9$ .
In all the panels the dotted line refers to $\NRF_{\pm}$ in the limit $\lambda \gg 1$.}\label{f:NRF}
\end{figure}
When we consider as input state
$| {\rm TWB} \rangle \rangle_{a_1,a_2}  | \sqrt{\mu}\,e^{i\psi} \rangle_{b_1} | \sqrt{\mu}\,e^{i\psi} \rangle_{b_2}$,
by using eqs.~(\ref{in:out:int}), (\ref{COHstat}) and (\ref{TWBstat}) we can easily
find:
\begin{eqnletter}
\label{NVN12}
\langle N_{k}\rangle &=&  \langle N\rangle^{\TWB}_{\eta\tau_{k}}
+\langle N\rangle^{\coh}_{\eta\tau_{k}} ,
\label{Ni} \\[1ex]
\langle\delta N_{k}^{2}\rangle &=& \, \langle\delta N^{2}\rangle^{\TWB}_{\eta\tau_{k}}
+\langle\delta N^{2}\rangle^{\coh}_{\eta\tau_{k}}
+ 2 \langle N\rangle^{\TWB}_{\eta\tau_{k}} \langle N\rangle^{\coh}_{\eta\tau_{k}} ,
\label{V(Ni)}\\[1ex]
\langle\delta N_{1} \delta N_{2}\rangle &=& \, \langle\delta N_{1} \delta N_{2}\rangle^{\TWB}_{\eta\tau_{1}\tau_{2}} 
- 2 \sqrt{\langle\delta N_{1} \delta N_{2}\rangle^{\TWB}_{\eta\tau_{1}\tau_{2}} \langle N\rangle^{\coh}_{\eta\tau_{1}} \langle N\rangle^{\coh}_{\eta\tau_{2}}}\, \cos(2 \psi )\,.
\label{cov(N1N2)}
\end{eqnletter}
It is interesting to note that the covariance $\langle\delta N_{1} \delta N_{2}\rangle$ depends
on the coherent fields phase $\psi$: if we set $\psi = \pi/2$ the covariance is maximised
whereas if $\psi = 0$, we can also has a negative value for $\langle\delta N_{1} \delta N_{2}\rangle$,
corresponding to an anticorrelation of photon numbers. We can now write explicitly the
expression of the noise reduction factor, namely:
\begin{eqnarray}
& &\NRF_{\pm} =\\
& &\hspace{0.5cm}\frac{\frac12 \langle\delta(N_{1}\pm N_{2})^{2}\rangle^{\TWB}_{\eta\tau}
+ 2 \langle N\rangle^{\coh}_{\eta\tau}
\left[
\frac12 +
\langle N\rangle^{\TWB}_{\eta\tau}
\mp  \sqrt{ \langle\delta N_{1} \delta N_{2}\rangle^{\TWB}_{\eta\tau}}\, \cos(2 \psi )
\right]}
{\langle N\rangle^{\TWB}_{\eta\tau} + \langle N\rangle^{\coh}_{\eta\tau} }.\nonumber
\end{eqnarray}
It is clear that the condition $\psi = \pi/2$ (optimising of the photon number correlation)
minimises $\NRF_{-}$, while $\psi = 0$ (optimising of the anticorrelation) minimises
$\NRF_{+}$: from now on we consider these optimising conditions when we refer to
the $\NRF_{\pm}$. In fig.~\ref{f:NRF} we plot the $\NRF_{\pm}$ as a function of
$\tau_1 = \tau_2 = \tau$ and different values of the other parameters.
\par
Since we are considering correlations, it is natural to identify two regimes. The first one correspond to the
regime studied in section~\ref{s:coupled:qgrav}, in which $\kappa \equiv \mu\left(1-\tau\right)/\tau\lambda \ll1$,
corresponding to
$\langle N\rangle^{\coh}_{\eta\tau}\ll \langle N\rangle^{\TWB}_{\eta\tau}$. In this case
the intensity arriving at the detectors is dominated by the twin beams and, as one may expect
from the results of section~\ref{s:coupled:qgrav}, $\NRF_{-}$
drastically decreases whereas $\NRF_{+}$ grows (see fig.~\ref{f:NRF} for $\tau \to 1$).
If we expand the noise reduction factor up to the first order in $1-\tau$ we have:
\begin{equation}
\NRF_{-} \approx \, 1- \eta + \eta (1-\tau) \left[
1 + 2\mu +\frac{\mu\left(1-2\sqrt{\lambda^2+\lambda}\right)}{\lambda}
\right],
\end{equation}
and
\begin{equation}
\NRF_{+} \approx  \, 1+ \eta (1+ 2 \lambda) + \eta (\tau - 1)\left[
1 + 2\lambda +\frac{\mu\left(1-2\sqrt{\lambda^2+\lambda}\right)}{\lambda}
\right],
\end{equation}
respectively, and we used, as mentioned above, the corresponding optimising
values of $\psi$. In the limit $\lambda \gg 1$ we also have
\begin{eqnletter}
\NRF_{-} &\approx& 1 - \eta \tau\,, \\[1ex]
\NRF_{+} &\approx& 1 + \eta \tau (1+2\lambda)\,.
\end{eqnletter}
Overall, this regime allows obtaining
a huge quantum enhancement in phase correlation measurement by exploiting the
twin-beam correlations also in the presence of a large classical power circulating
into the interferometer.
\par
In the other regime we have $\mu\left(1-\tau\right) \gg \lambda\tau$ and, thus,
$\langle N\rangle^{\coh}_{\eta\tau}\gg \langle N\rangle^{\TWB}_{\eta\tau}$. Now
(still considering the corresponding optimizing values of $\psi$) we find:
\begin{equation}
\NRF_{\pm} \approx 1 - 2\eta\tau\left[ \sqrt{\lambda(1+\lambda)} - \lambda\right] < 1,
\label{NRF+-(B)}
\end{equation}
 which, in the limit $\lambda \gg 1$ reduces to $\NRF_{\pm} \approx 1 - \eta \tau$ (see the dotted lines
 in fig.~\ref{f:NRF}). Since $\NRF_{\pm} < 1$, the number of photons are always correlated or
 anticorrelated beyond the classical limit. It is worth noting that this correlation can be
 also really bright, since, in real experiments, $\langle N\rangle^{\coh}_{\eta\tau}= \eta\mu (1-\tau)$
 can be very large \cite{coupled:15}.
\par
\begin{figure}[h!]
\begin{center}
\includegraphics[width=0.95\textwidth]{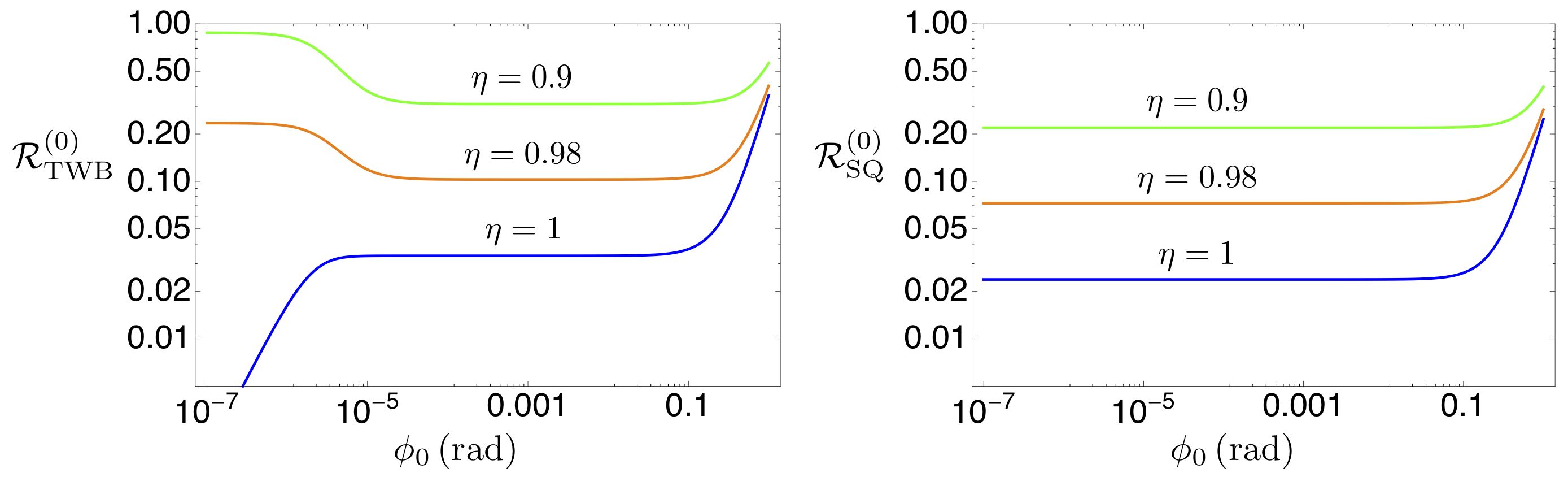}
\end{center}
\vspace{-0.5cm}
\caption{Log-log plot of the ratios ${\cal R}^{(0)}_{\rm TWB}$ (left) and ${\cal R}^{(0)}_{\rm SQ}$ (right)
as functions of the central phase $\phi_0$ measured by the interferometers and different
values of the detection efficiency $\eta$. When ${\cal R}^{(0)} < 1$
we have an advantage with respect to the classical case. We set $\lambda = 10$,
$\mu = 3 \times 10^{12}$ and $\psi = \pi/2$.}\label{f:ratio:eta}
\end{figure}
\begin{figure}[h!]
\begin{center}
\includegraphics[width=0.95\textwidth]{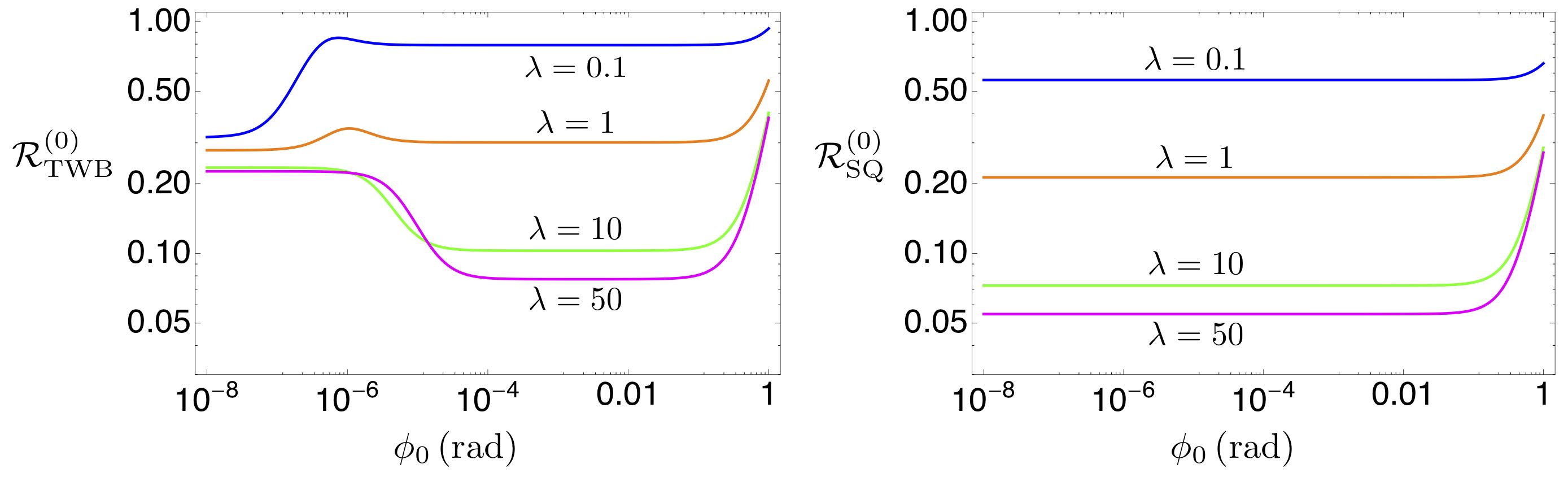}
\end{center}
\vspace{-0.5cm}
\caption{Log-log plot of the ratios ${\cal R}^{(0)}_{\rm TWB}$ (left) and ${\cal R}^{(0)}_{\rm SQ}$ (right)
as functions of the central phase $\phi_0$ measured by the interferometers and different
values of the detection efficiency $\lambda$. When ${\cal R}^{(0)} < 1$
we have an advantage with respect to the classical case. We set $\eta = 0.95$,
$\mu = 3 \times 10^{12}$ and $\psi = \pi/2$.}\label{f:ratio:lambda}
\end{figure}

We now focus our attention of the effect of small deviations from the optimal working regime
$\phi_0=\phi_{1,0}=\phi_{2,0}=0$
on the ratio ${\cal R}^{(0)}$ between the quantum and the classical uncertainties of the
covariance estimation introduced in the previous section (see fig.~\ref{f:ratio:unc}). Since
we are interested in having strong nonclassical correlations, we set $\psi = \pi/2$ and consider
the limit $\mu \gg 1$. In fact, a quantum strategy is supposed to enhance the sensitivity when
hight power is circulating in the interferometer. The results are plotted in figs.~\ref{f:ratio:eta} and \ref{f:ratio:lambda}
for both the twin beams and two independent squeezed vacua as inputs (and, of course, the coherent states).
\par
In the case of twin beams (left panels of figs.~\ref{f:ratio:eta} and \ref{f:ratio:lambda}) we can
in general identify two regions: one for $\phi_0 < 10^{-6}$ and the other for $10^{-5} < \phi_0 < 10^{-1}$.
These regions depend on the choice of the involved parameters and their relations.
\par
If $\phi_0$ is very small, the interferometer transmissivities are high and, thus, we have the
regime $\langle N\rangle^{\coh}_{\eta\tau}\ll \langle N\rangle^{\TWB}_{\eta\tau}$ in which
twin-beam correlations are predominant. This is the regime considered in the previous section
and in ref.~\cite{coupled:13}. In the second region, for $10^{-5} < \phi_0 < 10^{-1}$,
the transmissivity of the interferometers is such that the regime of bright quantum correlation
 $\langle N\rangle^{\coh}_{\eta\tau}\ll \langle N\rangle^{\TWB}_{\eta\tau}$ is reached: we find a flat
 ${\cal R}^{(0)}_{\rm TWB}$.
\par
When we consider two independent squeezed states instead of the twin beams
(see the right panels of figs.~\ref{f:ratio:eta} and \ref{f:ratio:lambda})
we can write the following expansions for $\mu \gg 1$ \cite{coupled:15}:
\begin{eqnletter}
\label{U-SQ(phi)}
\mathcal{R}_{\rm SQ}^{(0)} &\approx&
 1-\frac{\eta(1 +  \cos \phi_{0} )}{2}+\frac{\eta\cos^{2}(\phi_{0}/2)}{4 \lambda}\quad
\tx{if\ } \lambda\gg1, \\[1ex]
\mathcal{R}_{\rm SQ}^{(0)} &\approx&
 1-\eta  (1+\cos\phi_{0} ) \sqrt{\lambda }(1-\sqrt{\lambda })\quad
\tx{if\ } \lambda\ll1, \label{U-SQ(phi)b}
\end{eqnletter}
that are both less than 1 and, thus, show the advantage of using quantum light. From these
expansions and from the right panel of figs.~\ref{f:ratio:eta} and \ref{f:ratio:lambda}
we can see that the best results are obtained for $\phi_0$ very close to 0. Moreover,
if we consider the interval of phase values $10^{-5} < \phi_0 < 10^{-1}$
in the limit $\mu \gg 1$, we have ${\cal R}^{(0)}_{\rm TWB} \approx
 \sqrt{2}\,{\cal R}^{(0)}_{\rm SQ}$.
\par
\begin{figure}[h!]
\begin{center}
\includegraphics[width=0.95\textwidth]{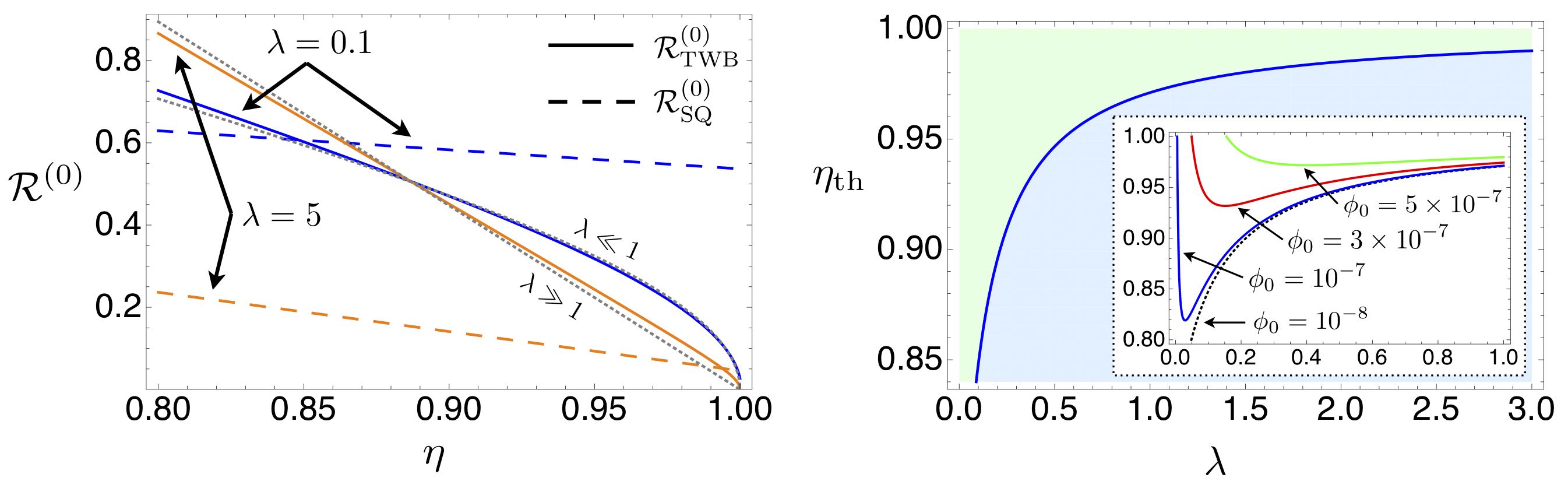}
\end{center}
\vspace{-0.5cm}
\caption{(Left) Plot of the ratio ${\cal R}^{(0)}_{\rm TWB}$ (solid lines) and ${\cal R}^{(0)}_{\rm SQ}$
(dashed lines) as a function of the detection efficiency $\eta$ for two different values of the average
number of photons per single mode $\lambda$ of the twin beams and the squeezed sates.
The grey dotted lines refer to ${\cal R}^{(0)}_{\rm TWB}$
in the limits $\lambda \gg 1$ and $\lambda \ll 1$.
(Right) Plot of the threshold value $\eta_{\rm th}$ as a function of $\lambda$: if
$\eta > \eta_{\rm th}$ then ${\cal R}^{(0)}_{\rm TWB} < {\cal R}^{(0)}_{\rm SQ}$ (upper
shaded region of the plot). The inset shows $\eta_{\rm th}$ as a function of $\lambda$
(we dropped the axes labels) for different values of $\phi_0$ (the dotted line is the
same of the main plot).
In both the panels we set  $\mu = 3 \times 10^{12}$, $\psi = \pi/2$ and $\phi_0 = 10^{-8}$.
}\label{f:versus:eta}
\end{figure}
In the left panel of fig.~\ref{f:versus:eta} we plot the ratio ${\cal R}^{(0)}$ for the twin beams (solid lines)
and the squeezed states (dashed lines) as functions of $\eta$ and a particular choice of the other involved
parameters. There exists a threshold value $\eta_{\rm th}$ of the quantum efficiency such that
for $\eta > \eta_{\rm th}$ the strategy based on twin beams outperforms the one exploiting independent
squeezed states (see fig.~\ref{f:versus:eta}, right panel).
\par
Overall, we can conclude that quantum light injected into the free ports of interferometers is a useful and
practical resource to improve the measurement of phase-correlation covariance. On the one hand,
the results obtained for two interferometers confirm the advantage of exploiting squeezed beams
and highly excited coherent states at the inputs.
This is the analogue of the case of single interferometers considered in
section~\ref{s:bounds}. On the other hand, the promising results concerning the twin beams show
that there is room for realistic applications of bipartite continuous-variable entanglement in
interesting and unexplored areas requiring high-precision correlation measurements, such as the
test of quantum gravity theories.


\section{Conclusion}\label{s:outro}

In these pages we have presented some of the theoretical and experimental results 
obtained by applying the tools of classical and quantum estimation theory to
optical systems.
Firstly, we have investigated the role of single- and two-mode
squeezing in active and passive interferometers, finding the regimes in which
quantum resources can improve the sensitivity of the interferometer also
in the presence of losses.
Then we have considered a quantum illumination protocol
where two quantum correlated beams are used to detect the presence
of an object embedded in a very noisy and predominant background. Also in
this case the theoretical and experimental results show that the use of
quantum light outperforms schemes based on classical correlated beams.
Eventually, our research was focused on interferometric schemes involving two
interferometers. This kind of setup has been recently proposed to test the
effects of the noncommutativity of the position and momentum operators
at the Planck scale predicted by a quantum gravity theory, giving birth to the
so-called holographic noise. We have theoretically
shown that coupling the two interferometers by means of two continuous-variable
entangled beams can improve the sensitivity to the holographic noise or,
more in general, to the correlation measurements.
\par
The quest for advanced schemes to measure a phase shift with high accuracy,
to detect the presence of an evanescent object in a noisy environment
or to unveil very faint and subtle phenomena that may confound with the background
is indeed amazing and always brings to
new exiting results (also thank to the technological advances).
The research we reviewed in this paper shows that there is
room for innovative-, quantum-enhanced schemes which can shed new light not only
on well-known areas, but also on still unexplored realms.


\acknowledgments
As any researcher knows, ``this is not just our work, but our dream and a big part of our life''.
The author would like to express his gratitude to all the scientists involved in the
``Light Correlations for High-precision Innovative Sensing -- LiCHIS''
project (funded by MIUR through the FIRB project nr.~RBFR10YQ3)
which has led to most of the results presented in these pages and, in particular,
to Alessia Allevi, Alice Meda, Matteo Bina, Ivano Ruo-Berchera, Ivo P.~Degiovanni,
Carlo Sparaciari, Maria Bondani, Marco Genovese and Matteo~G.~A.~Paris. Useful and stimulating
discussions with the people of the Quantum Technology Lab of the University of Milan
are also kindly acknowledged and, among them, with Claudia Benedetti, Fabrizio Castelli,
Simone Cialdi, Marco G.~Genoni, Nicola Piovella and Dario Tamascelli.

\appendix

\section{Beam splitter}\label{app:BS}
The beam splitter is a common device we can find in any quantum optics experiment.
The Hamiltonian describing the beam-splitter interaction can be written as
$\hH = g \ha^{\dag}\hb + g^{*} \ha\hb^{\dag}$, $g$ being the coupling constant,
and involves the two input bosonic field operators,  $\ha$ and $\hb$,
with commutation relations $[\ha,\ha^\dag] = {\mathbbm I}$ and
 $[\hb,\hb^\dag] = {\mathbbm I}$, respectively, and $[\ha,\hb]=0$.
For the sake of simplicity, we write the corresponding evolution operator as:
\begin{equation}\label{evol:BS}
\Ubs (\zeta)= \exp\left( \zeta\, \ha^{\dag}\hb - \zeta^{*}\,\ha\hb^{\dag} \right),
\end{equation}
where $\zeta= \phi\, \e^{i\theta}$.
To write the evolution operator in the form (\ref{evol:BS}) we applied the
transformation $\ha \to - i \ha$ which physically corresponds to impose a $\pi/2$ phase shift
to the mode $\ha$, namely $U_{\rm ph}^\dag(\pi/2)\,  \ha\, U_{\rm ph}(\pi/2) = -i \ha$, where
$U_{\rm ph}(\varphi) = \exp(-i\varphi\, \ha^{\dag}\ha)$ is the phase-shift operator and,
in general:
\begin{equation}
U_{\rm ph}^\dag(\varphi)\,  \ha\, U_{\rm ph}(\varphi) = \ha\, e^{-i\varphi}\,.
\end{equation}
\par
In order to calculate the Schr\"{o}dinger evolution of two states through a beam splitter
by applying $\Ubs (\zeta)$, we note that the two-boson operators
\begin{equation}
\hJ_{+} = \ha^{\dag}	\hb, \quad
\hJ_{-} = \ha\hb^{\dag},
\quad \mbox{and} \quad \hJ_{3} = \frac{1}{2}[\hJ_{+},\hJ_{-}] = \frac{1}{2}(\ha^{\dag}\ha - \hb^{\dag}\hb)\,,
\end{equation}
are a realisation of the $\mbox{SU}(2)$ algebra. Therefore, we can rewrite eq.~(\ref{evol:BS}) as follow:
\begin{eqnletter}
\Ubs (\zeta) &=&
\exp\left[\e^{i\theta}\tan\phi\, \ha^{\dag}\hb \right]\,
\left(\cos^2\phi\right)^{(\ha^{\dag}\ha - \hb^{\dag}\hb)/2}\,
\exp\left[-\e^{-i\theta}\tan\phi\, \ha\hb^{\dag} \right] , \\[1ex]
&=&
\exp\left[-\e^{-i\theta}\tan\phi\, \ha\hb^{\dag} \right]\,
\left(\cos^2\phi\right)^{-(\ha^{\dag}\ha - \hb^{\dag}\hb)/2}\,
\exp\left[\e^{i\theta}\tan\phi\, \ha^{\dag}\hb \right] ,
\end{eqnletter}
\par
Sometimes it is also useful to study the Heisenberg evolution of the input modes. In
this case we can use the identity:
\begin{equation}\label{thm:evol:Eq}
\e^{\hA}\, \hB\, \e^{-\hA} = \hB + \left[\hA,\hB\right] + 
\frac{1}{2!} \left[\hA,\left[\hA,\hB\right] \right] + 
\frac{1}{3!} \left[\hA,\left[\hA,\left[\hA,\hB\right]\right] \right] + \cdots
\end{equation}
which holds for two operators $\hA$ and $\hB$. In the presence of the beam splitter,
we obtain:
\begin{eqnletter}
\Ubs^{\dag}(\zeta)\, \ha\, \Ubs(\zeta) &=& \ha \cos\phi + \hb\, \e^{i\theta} \sin \phi ,
\\[1ex]
\Ubs^{\dag}(\zeta)\, \hb\, \Ubs(\zeta) &=& \hb \cos\phi - \ha\, \e^{-i\theta} \sin \phi ,
\end{eqnletter}
It is therefore clear that, due to the action of the beam splitter, each mode evolves into a linear
combination of the input modes and, for this reason, this kind of interaction is also called
\emph{two-mode mixing interaction}. The quantity $\tau = \cos^2 \phi$ is usually called
{\emph transmissivity} of the beam splitter.
\par
A {\it mirror} can be seen as a beam splitter with reflectivity equal to 1
Therefore, by setting $\phi = \pi/2$ and $\theta = 0$, we find the following mode transformations:
\begin{eqnletter}
\Ubs^{\dag}(\pi/2)\, \ha\, \Ubs(\pi/2) &=& \hb\ ,
\\[1ex]
\Ubs^{\dag}(\pi/2)\, \hb\, \Ubs(\pi/2) &=& - \ha\, .
\end{eqnletter}
\par
We note that the results described in this appendix depend
on the phase-shift transformation we initially applied to the mode $a$
to write the beam splitter evolution operator in the form (\ref{evol:BS}).

\section{Coherent states}\label{app:CS}

The coherent states are the closest approximation of the output state of a laser and
are the eigenvectors of the annihilation operator $a$, namely:
\begin{equation}
\ha \ket{\alpha} = \alpha \ket{\alpha},\quad \alpha\in {\mathbbm C}\,.
\end{equation}
Exploiting the completeness relation $\sum_n\ket{n}\bra{n} = {\mathbbm I}$ and the normalisation condition
$\braket{\alpha}{\alpha}=1$, we can find the photon number expansion of a coherent state,
which reads:
\begin{equation}
\ket{\alpha} = \e^{-|\alpha|^2/2} \sum_{n=0}^{\infty} \frac{\alpha^n}{\sqrt{n!}}\ket{n}\,,
\end{equation}
and the photon number distribution is:
\begin{equation}
p(n) = |\braket{n}{\alpha}|^2 = \e^{-|\alpha|^2}\,\frac{|\alpha|^{2n}}{n!}\,,
\end{equation}
that is a Poisson distribution with average number of photons
$N = \langle a^\dag a \rangle = |\alpha|^2$
and variance $\var[N] =  \langle (a^\dag a)^2 \rangle - \langle a^\dag a \rangle^2 =|\alpha|^2$.
\par
Coherent states can be obtained by applying the so-called displacement operator $\hD(\alpha)$ to
the vacuum state $| 0 \rangle$. The displacement operator is defined as:
\begin{equation}\label{displacement}
\hD(\alpha) = \exp\left( \alpha \ha^{\dag} - \alpha^* \ha \right).
\end{equation}
Given two operators $\hA$ and $\hB$ such that $\left[\hA,\hB\right] \in {\mathbbm C}$, we have:
\begin{eqnletter}
\label{BHC}
\exp\left(\hA+\hB\right) &=&
\exp\left(\hA\right) \exp\left(\hB\right) \exp\left\{-\frac12 \left[\hA,\hB\right]\right\},\\[1ex]
&=&
\exp\left(\hB\right) \exp\left(\hA\right) \exp\left\{\frac12 \left[\hA,\hB\right]\right\}.
\end{eqnletter}
If we use eqs.~(\ref{BHC}) to evaluate $\hD(\alpha) | 0 \rangle$ we have:
\begin{eqnletter}
\hD(\alpha)\ket{0} &=& \e^{-|\alpha|^2/2}\exp(\alpha\ha^{\dag})
\underbrace{\exp(\alpha^*\ha)\ket{0}}_{\displaystyle \ket{0}}\\[1ex]
&=& \e^{-|\alpha|^2/2}
\sum_{n=0}^{\infty} \frac{1}{n!}\left(\alpha \ha^{\dag}\right)^n
\ket{0}\\[1ex]
&=& \e^{-|\alpha|^2/2}
\sum_{n=0}^{\infty} \frac{\alpha^n}{\sqrt{n!}}\ket{n},
\end{eqnletter}
that is $\hD(\alpha)\ket{0} = \ket{\alpha}$.
\par
Other useful properties of the coherent states concerns the expectation values
of the quadrature operator
\begin{equation}
X_\theta = a \, e^{-i\theta} + a^{\dag} \, e^{i\theta}\,.
\end{equation}
Given the coherent state $| \alpha \rangle$, we have:
\begin{equation}
\langle X_\theta \rangle = \langle \alpha | X_\theta| \alpha \rangle = 2\, \Re{\rm e}[\alpha\, e^{-i\theta}]\,,
\end{equation}
for the first moment and
\begin{equation}
\var[ X_\theta ] = \langle X_\theta^2 \rangle - \langle X_\theta \rangle^2 = 1\,,
\end{equation}
for the variance which is independent of both the quadrature phase $\theta$ and
the coherent state amplitude $\alpha$. It is interesting to note that $\var[ X_\theta ]  = 1$ also for
$| \alpha \rangle = | 0 \rangle$, that is in the presence of the vacuum state: the value
$\var[ X_\theta ]  = 1$ is sometimes called the ``vacuum noise'' or the ``shot noise''
\footnote{It is worth noting that the actual value of $\var[ X_\theta ]$
depends on the definition of the quadrature $X_\theta$. In general, if we define
$X_\theta = \sigma_0 (a \, e^{-i\theta} + a^{\dag} \, e^{i\theta})$ we obtain, for a coherent state,
$\var[ X_\theta ] = \sigma_0^2$.}.
\par
Coherent states are \emph{minimum uncertainty} states. In fact, in general we have
\begin{equation}\label{ind:rel}
\var[X] \, \var[P] \ge 1\,,
\end{equation}
where $X = X_0$ and $P = X_{\pi/2}$. As a matter of fact, the inequality (\ref{ind:rel}) reaches the minimum
for coherent states, since $\var[X] = \var[P] = 1$. Nevertheless, there exists another class of minimum
uncertainty states, the squeezed states, which will be introduced in the following.

\section{Single--mode squeezed states}\label{app:SQ}

A state is called ``squeezed'' is the value of a quadrature variance is less than the vacuum state one
(in the present case less than 1).  The Hamiltonian associated with single-mode squeezing has the form
$\hH = g (\ha^{\dag})^2 + g^* \ha^2$ and the corresponding evolution operator can be written as:
\begin{equation}\label{sq:op}
\hS(\xi) = \exp\left[\frac12 \xi (\ha^{\dag})^2 -
\frac12 \xi^{*} \ha^2 \right],
\end{equation}
where $\xi = r\, \e^{i\psi}$. Upon introducing the operators
\begin{equation}
\hK_{+} = \frac12 (\ha^{\dag})^2, \quad 
\hK_{-} = \frac12 \ha^2, \quad
\hbox{and} \quad \hK_3 = - \frac12 [\hK_{+} ,\hK_{-} ] = \frac12 \left(\ha^{\dag}\ha + \frac12\right)\,,
\end{equation}
we obtain a boson realisation of $\mbox{SU}(1,1)$ algebra. Therefore we have the following
identity:
\begin{equation}\label{sq:s11}
\hS(\xi) = \exp\left[ \frac{\nu}{2\mu}(\ha^{\dag})^2\right]\,
\mu^{-(\ha^{\dag}\ha + \frac12)}\,
\exp\left[ -\frac{\nu^{*}}{2\mu}\ha^2\right],
\end{equation}
where $\mu = \cosh r$ and $\nu = \e^{i\psi} \sinh r$. It is straightforward to show that:
\begin{equation}\label{sq:mode:evol}
\hS^{\dag}(\xi)\,\ha\,\hS(\xi) = \mu \ha + \nu \ha^{\dag}\,,
\end{equation}
and if we apply the squeezing operator to the vacuum state, we obtain the
so called \emph{squeezed vacuum}, namely:
\begin{eqnletter}
\hS(\xi)\ket{0} &=& | 0, \xi \rangle\,,\\
&=& \frac{1}{\sqrt{\mu}} \sum_{n=0}^{\infty} \left(\frac{\nu}{2\mu}\right)^n
\frac{\sqrt{(2n)!}}{n!} \, \ket{2n}\,.
\end{eqnletter}
It is easy to show that the squeezed vacuum has the following relevant properties:
\begin{eqnletter}
\langle N \rangle &=& \sinh^2 r; \\[1ex]
\var[N] &=& 2 \sinh^2 r (\sinh^2 r + 1);\\[1ex]
\langle X_{\theta} \rangle &=& 0, \quad \forall \theta; \\[1ex]
\var[ X_{\theta}] &=& \e^{2r}\cos^2(\theta-\psi/2) + \e^{-2r}\sin^2(\theta-\psi/2).
\label{sq:var:quad}
\end{eqnletter}
where we set $\xi = r \, e^{i\psi}$. If we assume, without loss of generality,
a real squeezing parameter, {\it i.e.} $\xi = r \in {\mathbbm R}$, we have:
\begin{equation}
\var[ X] = e^{2r}\,, \var[ P] = e^{-2r} \Rightarrow \var[ X]\,  \var[ P] = 1\,,
\end{equation}
that is the state $| 0, r \rangle$ is a minimum uncertainty state where one of the quadrature,
$X$, has fluctuations larger than the vacuum one (for positive $r$) wile the other, $Y$, is larger.
The reader can prove that also the \emph{displaced squeezed state}:
\begin{equation}
| \alpha, \xi \rangle = \hD(\alpha) \hS(\xi) | 0 \rangle
\end{equation}
is still a minimum uncertainty state with quadrature variance given by Eq.~(\ref{sq:var:quad}) but with
$\langle X_\theta \rangle = 2\, \Re{\rm e}[\alpha\, e^{-i\theta}] \ne 0$ (note that $\langle X_\theta \rangle$
does not depend on the squeezing parameter $\xi$ but only on the displacement amplitude $\alpha$).

\section{Two-mode squeezed states}\label{app:TWB}

The Hamiltonian leading to two-mode squeezing is $\hH = g \ha^{\dag}\hb^{\dag} + g^* \ha\hb$,
which is similar to the single-mode squeezing one introduced in appendix~\ref{app:SQ},
but now it involves two different modes of the radiation field.
The evolution operator can be written as:
\begin{equation}\label{sq:2:op}
\hS_2(\xi) = \exp\left(\xi \ha^{\dag}\hb^{\dag} -\xi^{*} \ha\hb \right),
\end{equation}
where $\xi = r\, \e^{i\psi}$. We can obtain a realisation of $\mbox{SU}(1,1)$ algebra by
introducing the operators
\begin{equation}
\hK_{+} = \ha^{\dag} \hb^{\dag}, \quad
\hK_{-} = \ha \hb, \quad
\hbox{and} \quad\hK_3 = -\frac12 [\hK_{+} ,\hK_{-} ] = \frac12 (\ha^{\dag}\ha + \hb^{\dag}\hb + 1)\,.
\end{equation}
Therefore, as in the case of single-mode squeezing, we have:
\begin{equation}\label{sq:2:s11}
\hS_2(\xi) = \exp\left( \frac{\nu}{\mu}\ha^{\dag}\hb^{\dag}\right)\,
\mu^{-(\ha^{\dag}\ha + \hb^{\dag}\hb + 1)/2}\,
\exp\left( -\frac{\nu^{*}}{\mu}\ha\hb\right),
\end{equation}
where $\mu = \cosh r$ and $\nu = \e^{i\psi} \sinh r$. Under the action of $\hS_2(\xi)$ the
field operators $\ha$ and $\hb$ transforms as follows:
\begin{equation}\label{two:sq:mode:evol}
\hS^{\dag}_2(\xi)\,\ha\,\hS_2(\xi) = \mu \ha + \nu \hb^{\dag} \quad
\mbox{and} \quad
\hS^{\dag}_2(\xi)\,\hb\,\hS_2(\xi) = \mu \hb + \nu^* \ha^{\dag}.
\end{equation}
In analogy to the squeezed vacuum state, if we apply the two mode squeezing operator
to the vacuum state we obtain the \emph{two-mode squeezed vacuum}, namely:
\begin{equation}\label{TWB}
\hS_2(\xi)\ket{0} = \frac{1}{\sqrt{\mu}} \sum_{n=0}^{\infty} \left(\frac{\nu}{\mu}\right)^n
 \, \ket{n}\ket{n},
\end{equation}
or, if we introduce the parameter $\lambda = \e^{i\psi} \tanh r$:
\begin{equation}
\hS_2(\xi)\ket{0} = \sqrt{1-|\lambda|^2} \sum_{n=0}^{\infty} \lambda^n
 \, \ket{n}\ket{n},
\end{equation}
which is also referred to as twin-beam state (TWB), since a measurement of the photon number
on the two beams always leads to the same result.  Note that
\begin{equation}
|\lambda|^2 = \frac{N}{N+1}\,,
\end{equation}
where $N = \sinh^2 |\xi|$ is the average number of photon per mode.
\par
Remarkably, if we consider a single beam of a TWB, it is a \emph{thermal state}
with average number of photons $N$ and, thus, it is described by the density
operator:
\begin{equation}\label{thermal}
\hrho(N) = \frac{1}{1+N}
\sum_{n=0}^{\infty} \left( \frac{N}{1+N} \right)^n \ket{n}\bra{n}\,,
\end{equation}
\par
The states of the form given in eq.~(\ref{thermal}) are called thermal states because they have the
same analytical expression of the state describing a radiation of frequency $\omega$
at equilibrium at temperature $T$. In this last case, the average number of photons is:
\begin{equation}
N_{\rmpl{th}} = \frac{1}{\displaystyle e^{\hbar\omega/(k_{\rm B}T)} - 1}\,,
\end{equation}
where $k_{\rm B}$ the Boltzmann constant. It is also interesting to note that since
the reduced state of the TWB is a thermal state, it exhibits the maximum von Neumann
entropy for a fixed energy: being the TWB a pure state, this means that it is a maximally
entangled state.

\section{Bernoulli sampling from non-unit efficiency photodetection}\label{app:eta}

A photon-number-resolving detector allows to directly measure the photon number
distribution
\begin{equation}
p(n) = \bra{n} \hrho\ket{n}
\end{equation}
of an input state $\hrho$ and it is described
by the projectors $\ket{n}\bra{n}$ onto the photon number basis $\{ \ket{n} \}$,
with $n \in {\mathbbm N}$.
\par
\begin{figure}[h!]
\begin{center}
    \includegraphics[width=0.35\textwidth]{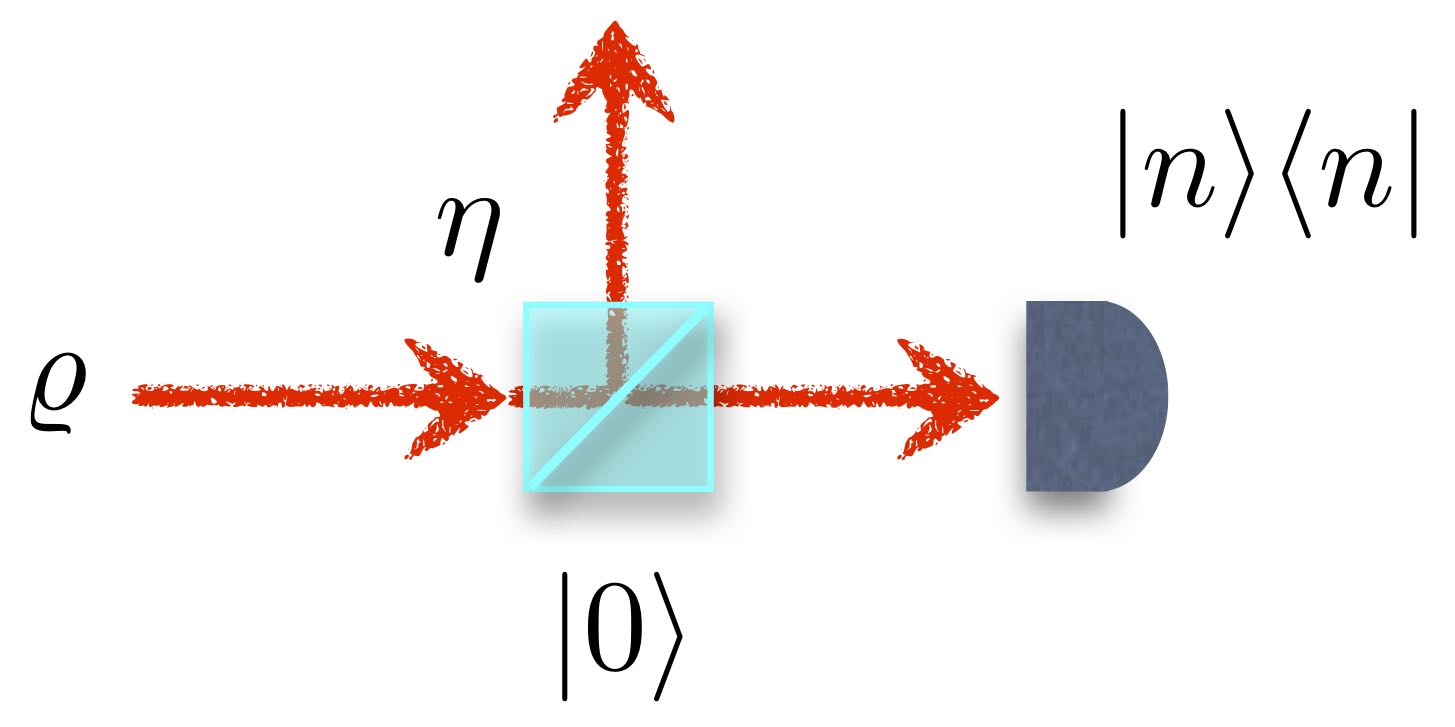}
\end{center}
\vspace{-0.5cm}   
\caption{\label{f:detection} A photon number resolving detector with quantum efficiency $\eta$ can be represented as an ideal photodetector and a beam splitter with transmissivity $\eta$ in front of it.
Note the presence of the vacuum state $\ket{0}$.}
\end{figure}

However, a realistic detector has a non-unit quantum efficiency $\eta$,
that can be seen as an overall loss of photons during the detection process.
From the theoretical point of view, a real photodetector can be modelled as a
beam splitter with transmissivity $\eta$ followed by an ideal
photon-number-resolving detector, as sketched in fig.~\ref{f:detection}.
In this scheme, before the detection the input state is mixed with the vacuum
state at the beam splitter and part of its photons is thus reflected and lost.
\par
If we send a single photon state $\ket{1}$ to the realistic detector, $\eta$
corresponds to the probability of detection. What happens when we send a\
Fock state $\ket{n}$, $n>1$? Starting from the model of
fig.~\ref{f:detection} and by using the results of appendix~\ref{app:BS}, we can explicitly
calculate the evolution of the two-mode input state $\ket{n}\ket{0}$
just after the beam splitter, described by the unitary operator $\Ubs(\eta)$, namely:
\begin{eqnletter}
\Ubs(\eta) \ket{n}\ket{0} &=& \frac{1}{\sqrt{n!}} \left(\sqrt{\eta}\, \ha^{\dag} -
\sqrt{1-\eta}\, \hb^{\dag}\right)^n \ket{0}\\
&=& \frac{1}{\sqrt{n!}} \sum_{k=0}^{n}{n \choose k} (-1)^{k} \sqrt{\eta^{n-k} (1-\eta)^{k}} \,
(\ha^{\dag})^{n-k}(\hb^{\dag})^k \ket{0}\\
&=& \sum_{k=0}^{n}(-1)^{k} \sqrt{{l \choose k} \eta^{n-k} (1-\eta)^{k}} \,
\ket{l-k}\ket{k}\\
&=& \sum_{m=0}^{n}(-1)^{h-m} \sqrt{{n \choose m} \eta^{m} (1-\eta)^{n-m}} \,
\ket{m}\ket{n-m}\,,
\end{eqnletter}
where we assumed that the states $\ket{n}$ and $\ket{0}$ refer to the modes
described by the bosonic filed operators $a$ and $b$, respectively.
Therefore, the probability $P(m;\eta)$ to detect $m$ photons, $m\le n$, is given by:
\begin{equation}
P(m;\eta) = {n \choose m} \eta^{m} (1-\eta)^{n-m}\,.
\end{equation}
Of course, if $\eta \to 1$ we have $P(m; \eta) \to p(m) = \delta_{l,m}$.
\par
It is now clear that if we know the \emph{actual} photon number statistics $p(n)$ of the state $\hrho$, then the \emph{detected} photon number statistics is:
\begin{eqnletter}
P(m;\eta) &=& \sum_{n=0}^{\infty} {n \choose m} \eta^{m} (1-\eta)^{n-m} p(n)\\
&=& \sum_{n=0}^{\infty} {n \choose m} \eta^{m} (1-\eta)^{n-m} \bra{n} \hrho\ket{n}\,.
\end{eqnletter}
\par
Exploiting this simple model based on the beam splitter, it is also straightforward to find the
evolution of a quantum state $\varrho$ through a dissipative channel: in this case the quantum
efficiency should be replaced by the overall loss parameter.


\end{document}